\documentclass[preprint,showpacs,preprintnumbers,superscriptaddress,amsmath,amssymb]{revtex4}

\usepackage{epsfig}
\usepackage{dcolumn}
\usepackage{bm}

\begin{document}

\newcommand{\wigner}[6]
{{
\left(
\begin{array}{ccc}
#1 & #2 & #3 \\
#4 & #5 & #6 \\
\end{array}
\right)
}}

\title{Search for non-Gaussianity in pixel, harmonic and wavelet space: compared and combined}

\author{Paolo Cabella}
\email{Paolo.Cabella@roma2.infn.it}
\affiliation{Dipartimento di Fisica, Universit\`a di Roma `Tor Vergata', Via della Ricerca Scientifica 1, I-00133 Roma, Italy}
\author{Frode Hansen}
\email{frodekh@roma2.infn.it}
\affiliation{Dipartimento di Fisica, Universit\`a di Roma `Tor Vergata', Via della Ricerca Scientifica 1, I-00133 Roma, Italy}
\author{Domenico Marinucci}
\email{marinucc@mat.uniroma2.it}
\affiliation{Dipartimento di Matematica, Universit\`a di Roma `Tor Vergata', Via della Ricerca Scientifica 1, I-00133 Roma, Italy}
\author{Daniele Pagano}
\affiliation{Dipartimento di Fisica, Universit\`a di Roma `Tor Vergata', Via della Ricerca Scientifica 1, I-00133 Roma, Italy}
\author{Nicola Vittorio}
\affiliation{Dipartimento di Fisica, Universit\`a di Roma `Tor Vergata', Via della Ricerca Scientifica 1, I-00133 Roma, Italy}

\date{\today}

\begin{abstract}
We present a comparison between three approaches to test non-Gaussianity of cosmic microwave background data. The Minkowski functionals, the empirical process method and the skewness of wavelet coefficients are applied to maps generated from non-standard inflationary models and to Gaussian maps with point sources included. We discuss the different power of the pixel, harmonic and wavelet space methods on these simulated almost full-sky data (with Planck like noise). We also suggest a new procedure consisting of a combination of statistics in pixel, harmonic and wavelet space.
\end{abstract}

\pacs{02.50.Ng, 95.75.Pq, 02.50.Tt, 98.80.Es}

\maketitle





\section{Introduction}

The fluctuations of the cosmic microwave background (CMB) are expected to be close to Gaussian distributed. In view of the increasing quantity of CMB experiments, it is now possible to check this assumption on data with growing resolution and sky coverage. Most models for the early universe predict some small deviations from Gaussianity; non-standard models of inflation \cite{nongi1,nongi2,nongi3,sabino,nongi5,nongi6,michele}, cosmic strings (See Ref.\cite{strings} for a review) and point sources. Detecting these small deviations would be of great importance for the understanding of the physics of the early universe. Also systematic effects like a non-symmetric beam and noise could give rise to non-Gaussian features. For this reason a non-Gaussianity check could reveal whether the impact of the instrumental effects on the data of the experiment is well understood.\\

The methods to search for non-Gaussianity in the literature mainly concentrate on implementing the test in three different spaces: (1) In pixel space: the Minkowski functionals \cite{novikov,gott} (which were used to set limits on the non-Gaussianity in the WMAP data \cite{WMAPKomatsu}), temperature correlation functions \cite{eriksen}, the peak to peak correlation function \cite{heavens}, skewness and kurtosis of the temperature field \cite{vittorio} and curvature properties \cite{dore,barreiro2}, to mention a few. (2) In harmonic space: analysis of the bispectrum and its normalized version \cite{phillips,komatsu,grazia,komatsuspergel} and the bispectrum in the flat sky approximation \cite{win}. The explicit form of the trispectrum for CMB data was derived in \cite{hu,kunz}. Phase mapping \cite{naselsky}. Applications to COBE, Maxima and Boomerang data have also drawn enormous attention and raised wide debate \cite{boom,polenta,cobeng1,cobeng2}. The empirical process method \cite{paper1,paper2,mar}. Finally, (3) wavelet space: \cite{barreiro1,barreiro3,mg,cobewav}. Traditionally, these tests are performed separately in each space. In this article, we will take methods in pixel- (the Minkowski functionals), harmonic- (the empirical process) and wavelet-space (skewness), and we will make a comparison for two different models of non-Gaussianity. We will also combine the methods in order to improve the total power. It should be noted that all the procedures we consider are non-parametric, that is they do not assume any a priori knowledge about the nature of non-Gaussianity.\\

We will use these methods on 100 maps generated from a non-standard inflationary model \cite{sabino} and on 100 maps where we have included point sources. We assess the performance of the methods in the different spaces for the different types of non-Gaussianity. We also propose a combined test which turns out to be more robust.\\

In section \ref{sect:mf}, we review the method of Minkowski functionals, in section \ref{sect:mfimpl} we describe our implementation of the method and in section \ref{sect:pixel} we define our proposed statistic. In section \ref{sect:ep}, we review the empirical process method while section \ref{sect:wav} is devoted to the wavelets. In section \ref{sect:comp} the methods are compared and applied to non-standard inflationary models, in section \ref{sect:point} to maps with point sources. Finally in section \ref{sect:concl} we summarize and comment on our results.

\section{Minkowski functionals}
\label{sect:mf}

To analyze a spherical map in terms of Minkowski functionals, we consider
the excursion sets, that is, the map subsets which exceed a given threshold
value. The threshold is labelled $\nu ,$ and it is treated as an independent
variable, on which these functionals depend. More precisely, considering the
normalized random field of temperature fluctuations, $u=\Delta T/\sigma
(\Delta T);$ we can define the 'hot region' Q as the ensemble of pixels $u_{i}$ higher than
the $\nu $ level : 
\begin{equation}
Q\equiv Q(\nu )=\{i|u(\theta _{i},\varphi _{i})>\nu \}\text{ .}
\end{equation}%
The three functionals of interest then are, up to constant factors
\cite{Minkowski}:

1) \emph{Area:} $M_{0}(\nu )$ is the total area of all hot regions.

2) \emph{Boundary length:} $M_{1}(\nu)$ is proportional to the total length
of the boundary between cold and hot regions

3) \emph{Euler characteristic or genus: }$M_{2}(\nu )$, a purely topological
quantity, counts the number of isolated hot regions minus the number of
isolated cold regions, i.e. the number of connected components in Q minus the number of 'holes'.

The rationale behind these statistics can be explained from mathematical
results in Hadwiger (1959); in particular, these results can be interpreted
by stating that all the morphological information of a convex body is
contained in the Minkowski functionals (Winitzki and Kosowsky, 1997)); here,
by morphological we mean the properties which are invariant under
translations and rotations and which are additive \cite{Tomita,Worsley,Schmalzing}. The
three statistics, normalized by the area density, can then be expressed as 
\begin{eqnarray}
M_{0}(\nu ) &=&\frac{1}{A}\int_{Q}dA  \label{m2} \\
M_{1}(\nu ) &=&\frac{1}{4A}\int_{\partial Q}dl \\
M_{2}(\nu ) &=&\frac{1}{2\pi A}\int_{\partial Q}\kappa dl
\end{eqnarray}%
where $\partial Q$ is the contour of the region $Q$; $dA$ and $dl$ are the
differential elements of $Q$ and $\partial Q$, respectively; $\kappa $ is
the geodetic curvature of $dl$ .

The expected values for a given thresholds depends on a single parameter $%
\tau$ given for a Gaussian field by \cite{Tomita}: 
\begin{eqnarray}  \label{m2_teor}
M_{0}(\nu)&=&\frac{1}{2}\left[1-erf\left(\frac{\nu}{\sqrt(2)}\right)\right]
\\
M_{1}(\nu)&=&\frac{\sqrt{\tau}}{8} \ exp\left(-\frac{\nu^2}{2}\right) \\
M_{2}(\nu)&=&\frac{\tau}{\sqrt{8\pi^3}}\nu \ exp\left(-\frac{\nu^2}{2}\right)
\end{eqnarray}
with : 
\begin{equation}  \label{eq:tau_der}
\tau=\frac{1}{2}\langle u_{;i} \ u_{;i} \rangle
\end{equation}
where semicolon indicates the covariant derivative on the sphere.

In the case of CMB (\ref{eq:tau_der}) reduces to \cite{Schmalzing}: 
\begin{equation}
\tau =\sum_{l=1}^{\infty }(2l+1)C_{\ell }\frac{l(l+1)}{2}  \label{eq:tau_cl}
\end{equation}%
where $C_{\ell }$ is the angular power spectrum.

An immediate consequence of the above formulae is that, although the
expected value of the first Minkowski functional is invariant with respect
to the dependence structure of $\Delta T$, for the second and third Minkowski
functional this is not the case and calibration for a given angular power
spectrum $C_{\ell}$ is needed. Moreover, even for the first Minkowski
functional, knowledge of the angular power spectrum is required for a Monte
Carlo evaluation of its variance. This can be viewed as a drawback,
and because of this some effort has been undertaken to provide at least some
crude upper bound for the functionals' variance (Winitzki and Kosowsky
(1997)).

\section{The Implementation of Minkowski Functionals}
\label{sect:mfimpl}

\begin{figure}[tbp]
\caption{Comparison between Minkowski functionals computed in the tangent
plane approximation and their analytical predictions (solid line) see Eqs.(\ref{m2_teor}). Squares refer to tangent
planes of size $\simeq (12^0\times 12^0)$, triangles refer to tangent planes
of size $\simeq (24^0\times 24^0)$. }
\label{fig:Mink_teor_mc}
\begin{center}
\leavevmode \epsfig {file=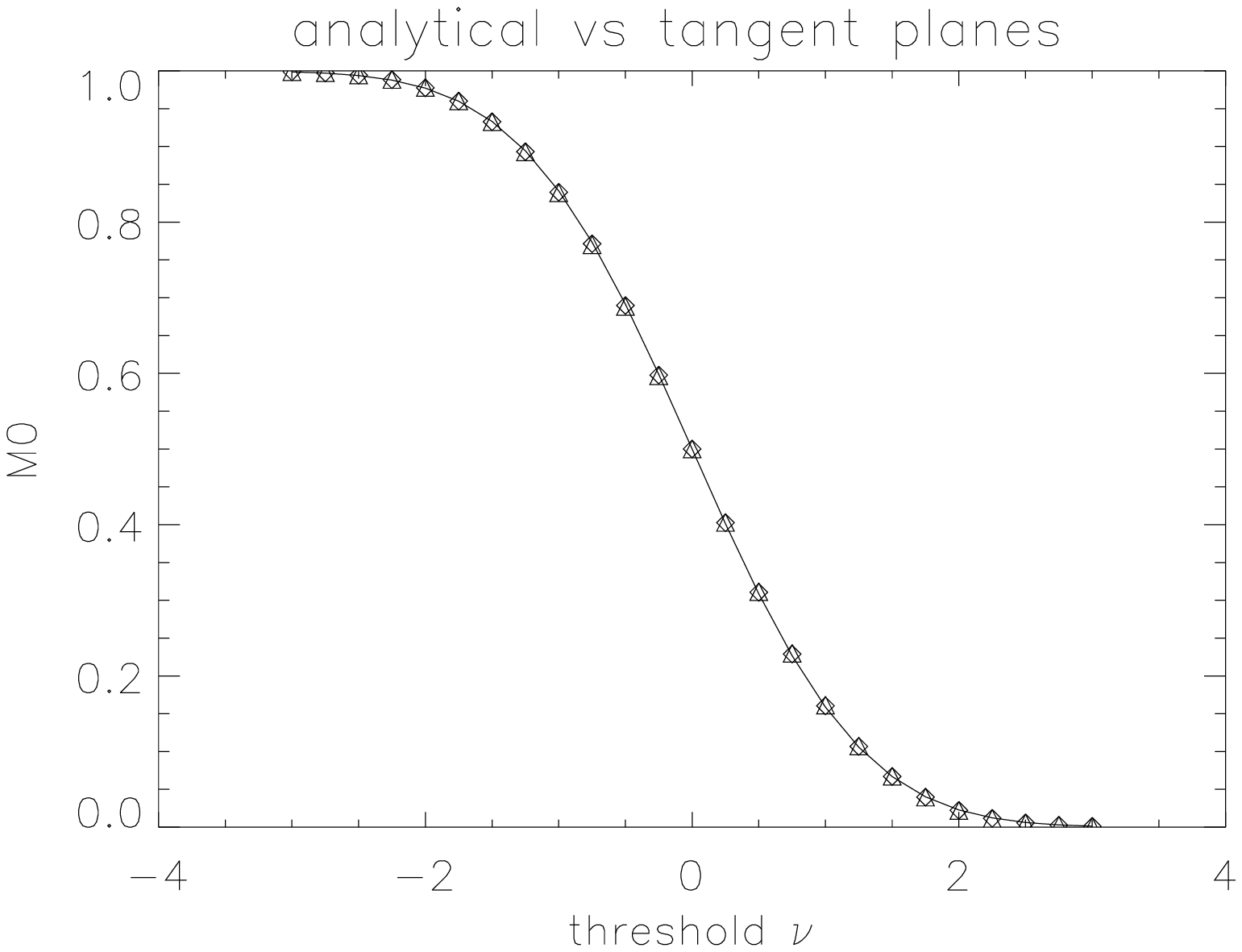,width=7cm,height=7cm} \epsfig {%
file=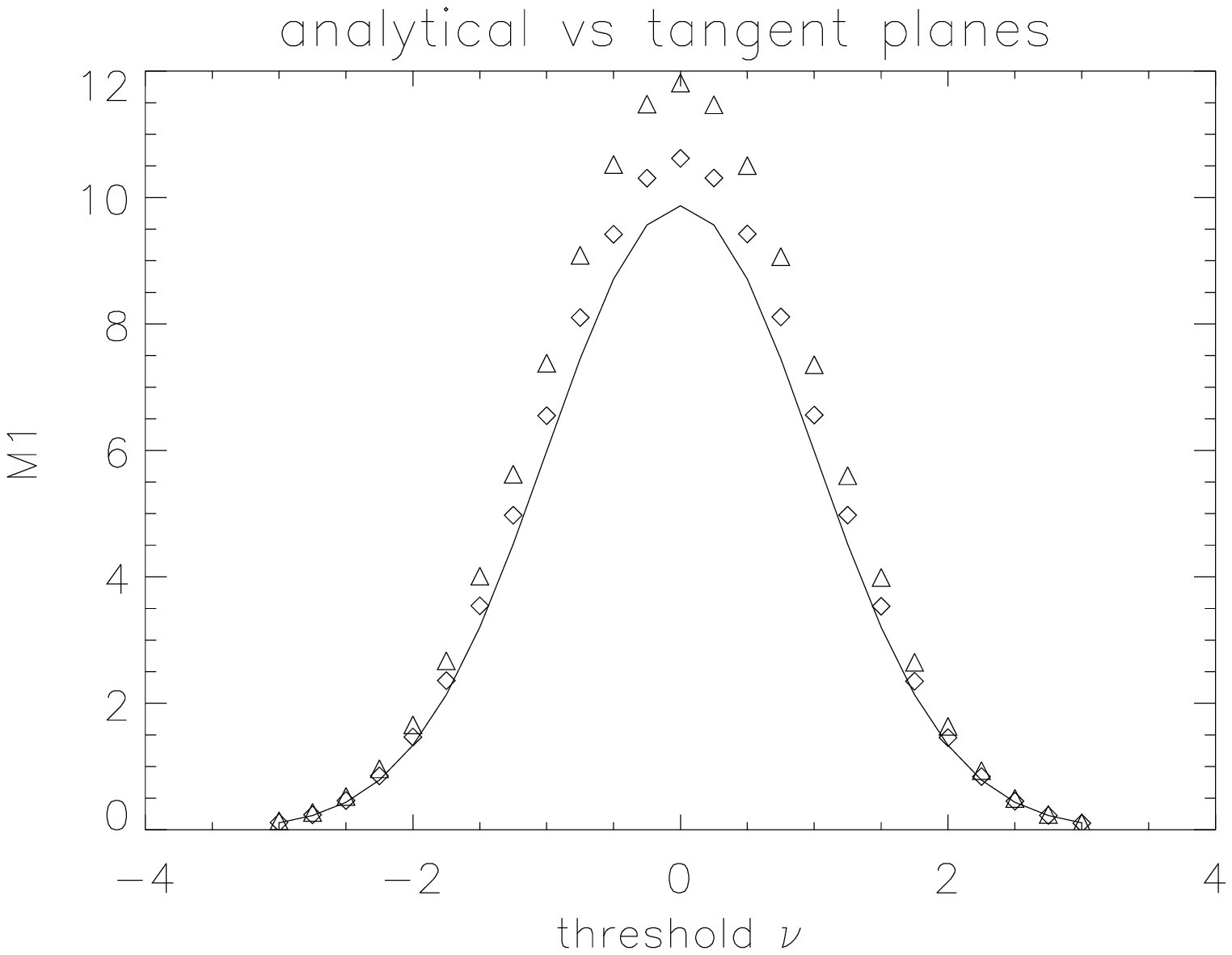,width=7cm,height=7cm} \epsfig {%
file=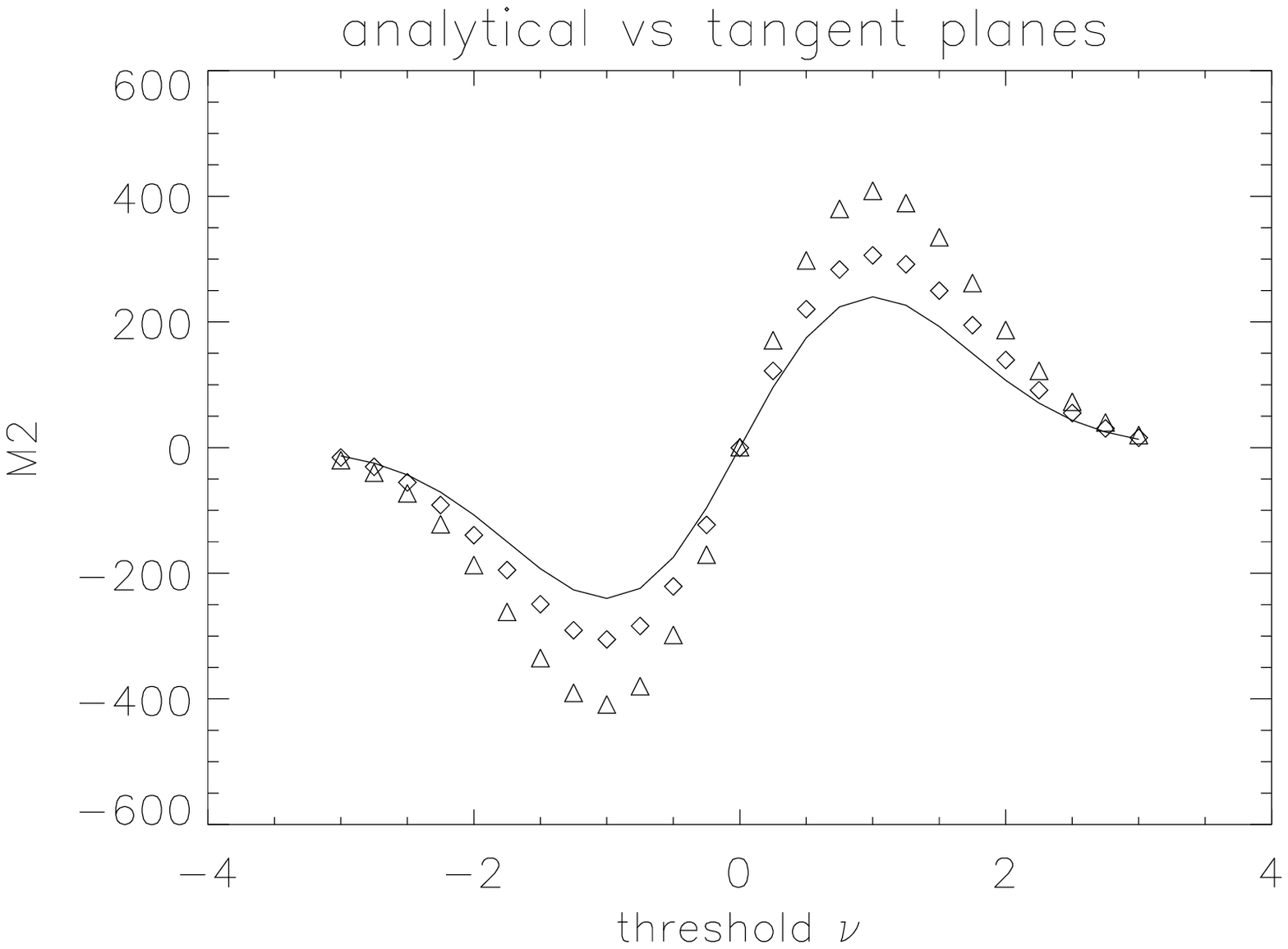,width=7cm,height=7cm}
\par

\end{center}
\end{figure}

\begin{figure}[tbp]
\begin{center}
\leavevmode \epsfig {file=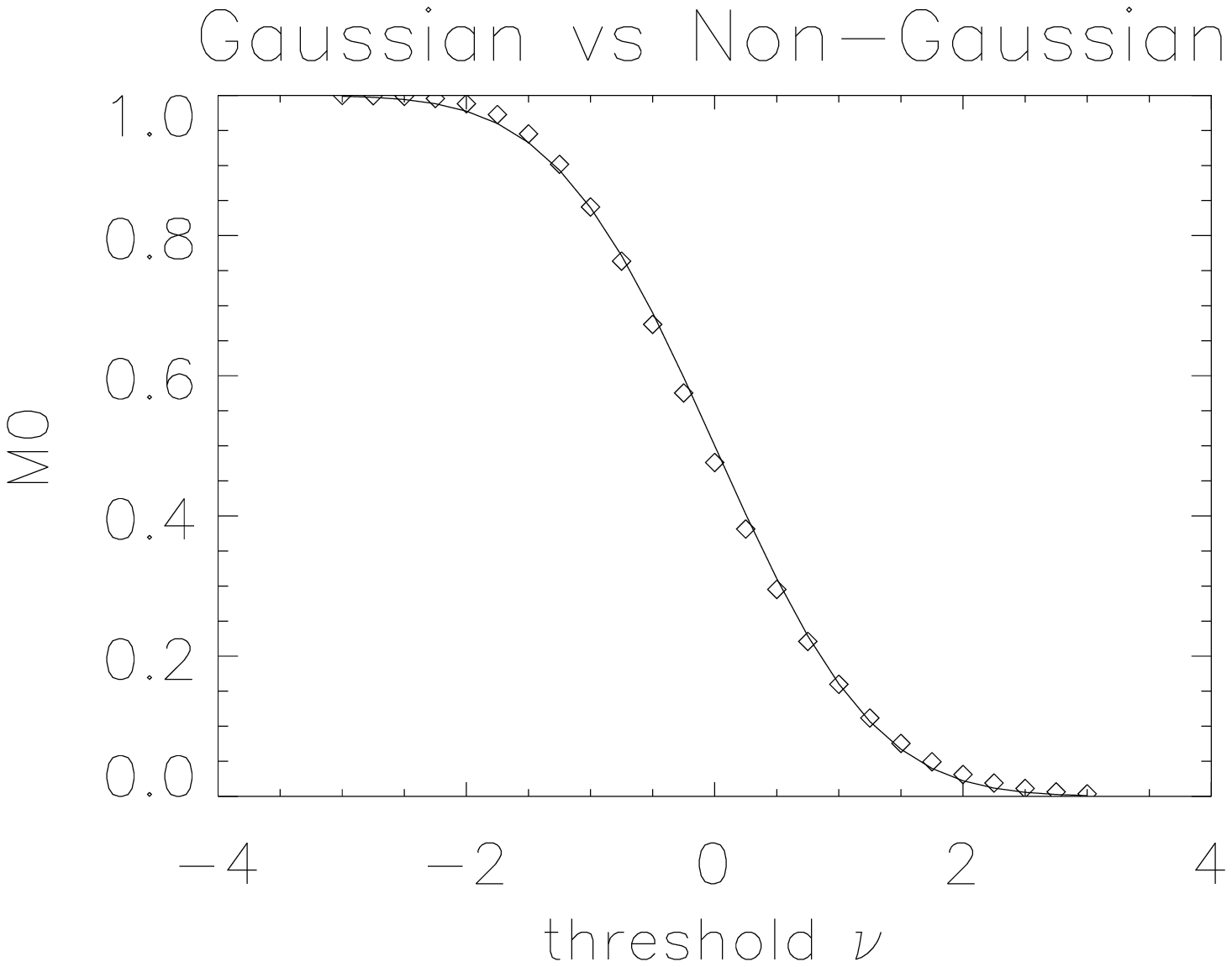,width=7cm,height=7cm} \epsfig {%
file=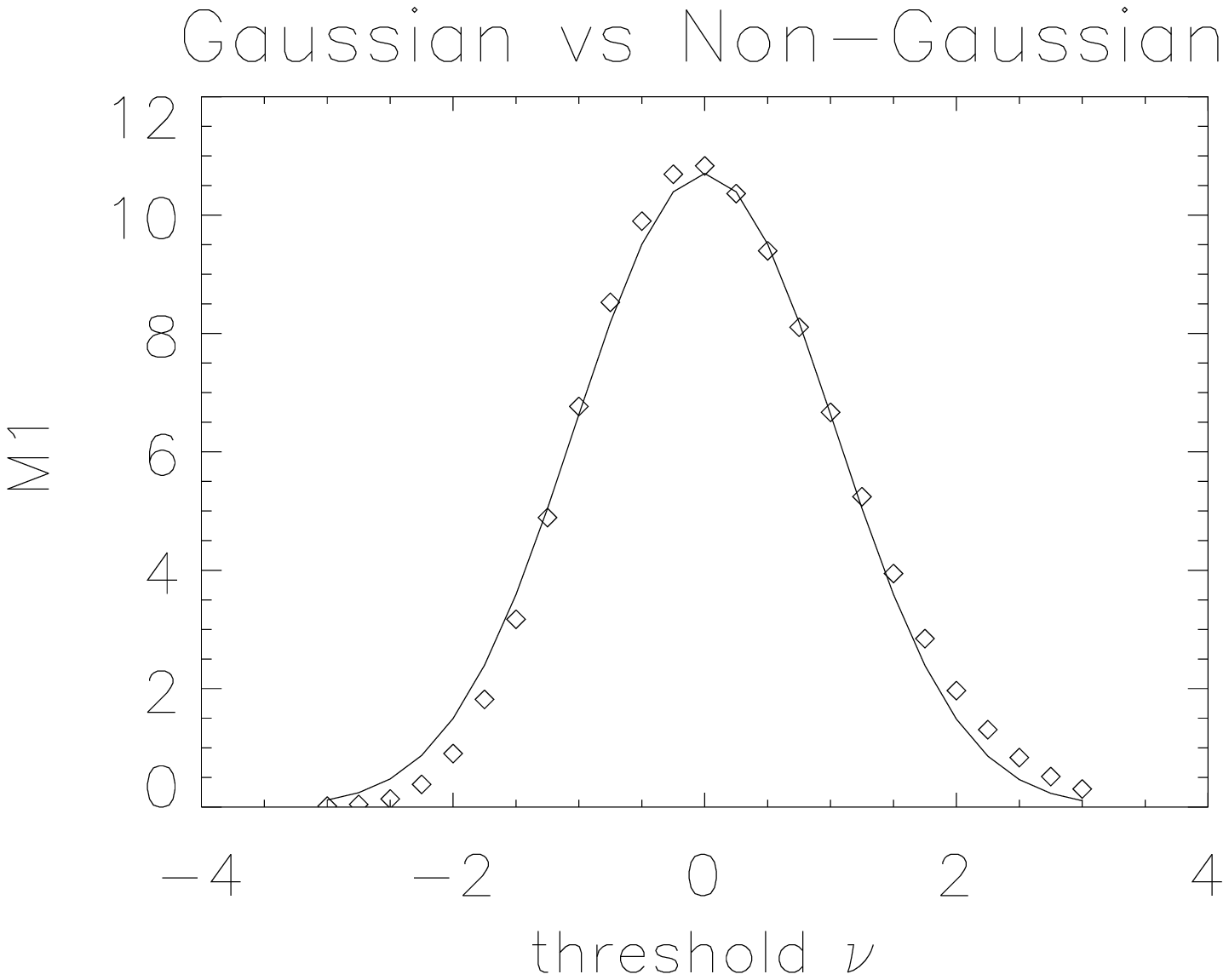,width=7cm,height=7cm} \epsfig {%
file=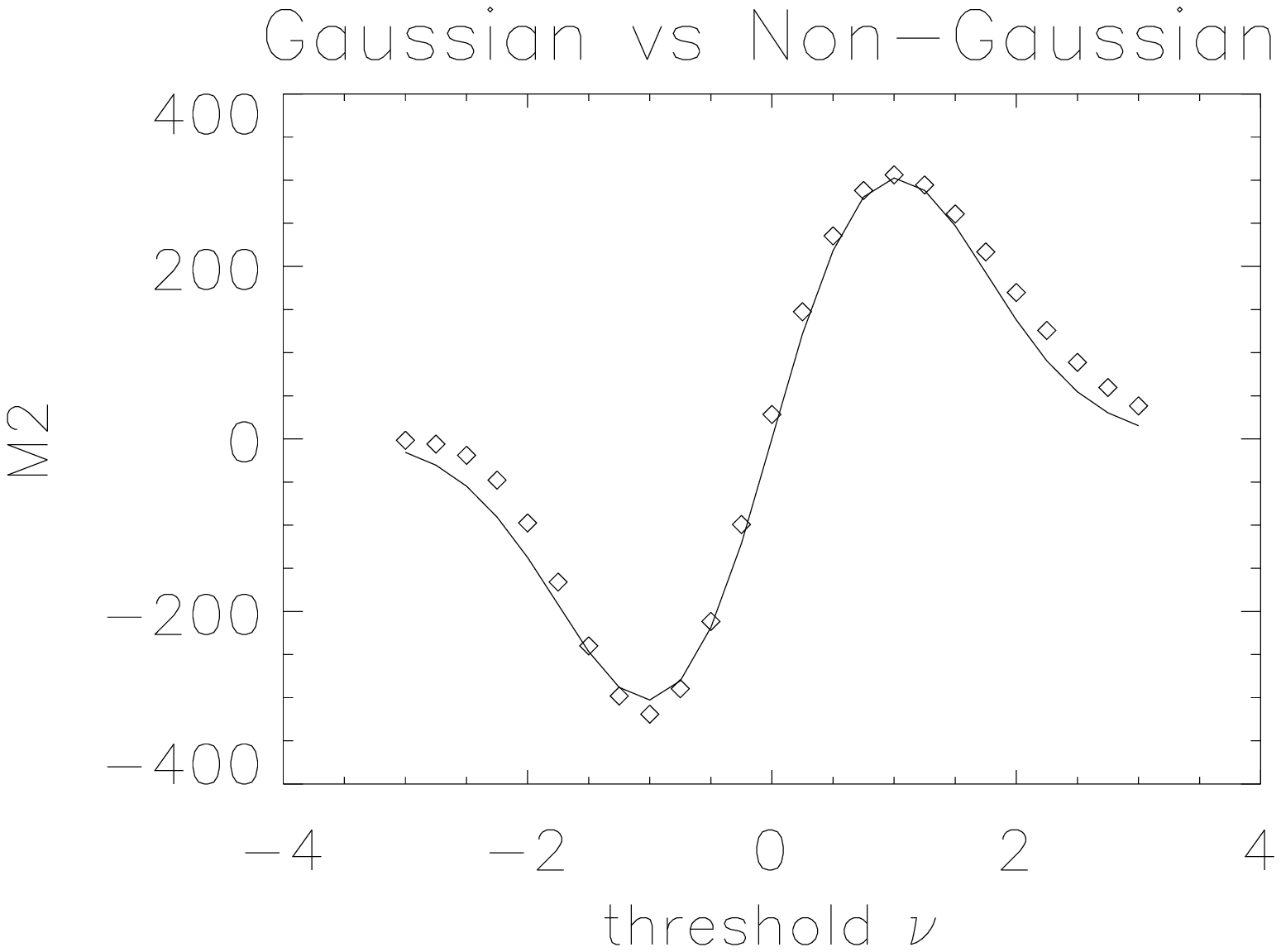,width=7cm,height=7cm}
\par

\end{center}
\caption{A comparison between a Gaussian map (solid line) and non-Gaussian
map with $f_{NL}=1000$ (squares) (the $f_{NL}$ factor will be explained in section \ref{sect:comp}).}
\label{fig:Mink_fnl1000_mc}
\end{figure}

In order to estimate the three functionals we simulate a map of the CMB \cite{healpix} with a known power spectrum, and then we cut the maximum number of
independent tangent planes of dimension $\simeq $ $12^{\circ }\times
12^{\circ }$; in this way it is easy to calculate the values of the three
Minkowski functionals in the flat-sky limit, taking into account the possibility
of gaps (galactic cut, polar calottes). Due to projection effects, finite pixel size and the dimension of the tangent planes, we find a deviation of the simulated values with respect to the analytical spherical predictions (eqs.\ref{m2_teor}). Note however that the shape of the curves
is unaffected, which is not the case when non-Gaussianities are present (see
fig. \ref{fig:Mink_fnl1000_mc}). In fig. \ref{fig:Mink_teor_mc} we show a comparison between analytical expectation values and the values computed on the tangent planes.

\section{Test of non-Gaussianity in pixel space}
\label{sect:pixel}

In order to test non-Gaussianity we use a test defined by:

\begin{equation}  \label{eq:quantile}
I_i=\int |M^i(\nu)-\bar{M}^i(\nu)|d\nu \:\:\:\:i\:=\:0,\:1,\:2
\end{equation}

Our procedure is as follows;

\begin{figure}[tbp]
\leavevmode
\par
\begin{center}
\leavevmode \epsfig {file=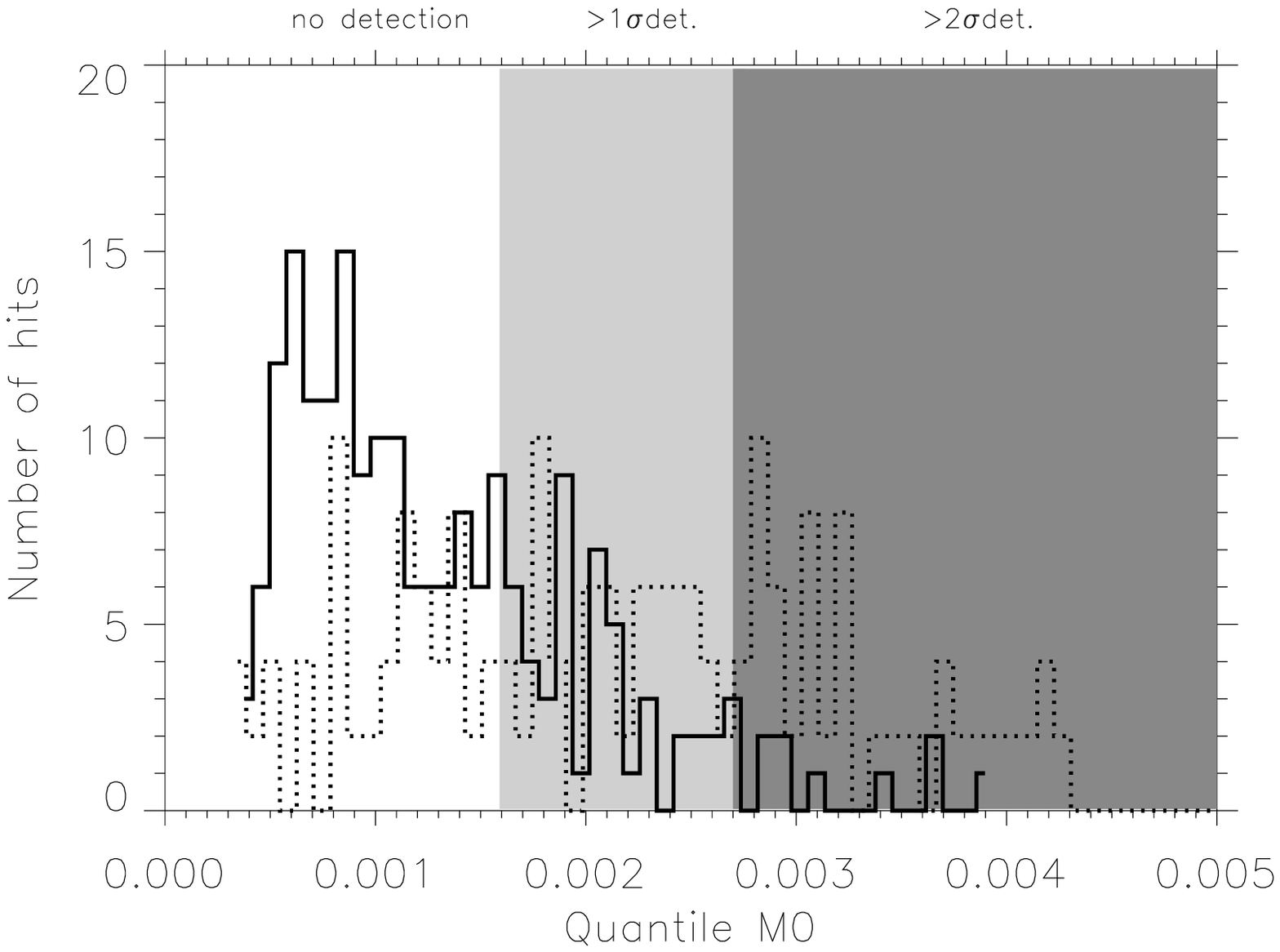,width=7cm,height=7cm} \epsfig {%
file=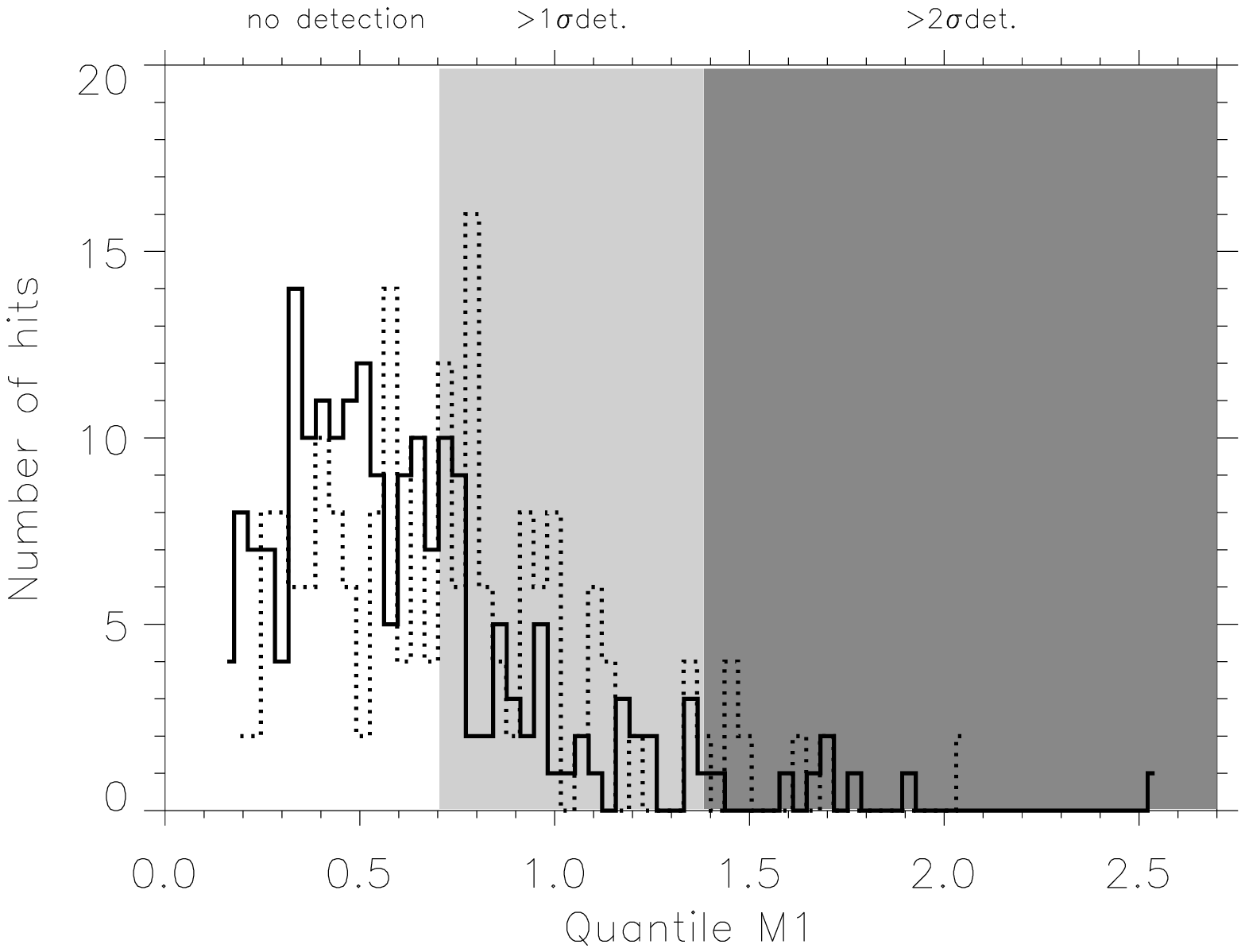,width=7cm,height=7cm} \epsfig {%
file=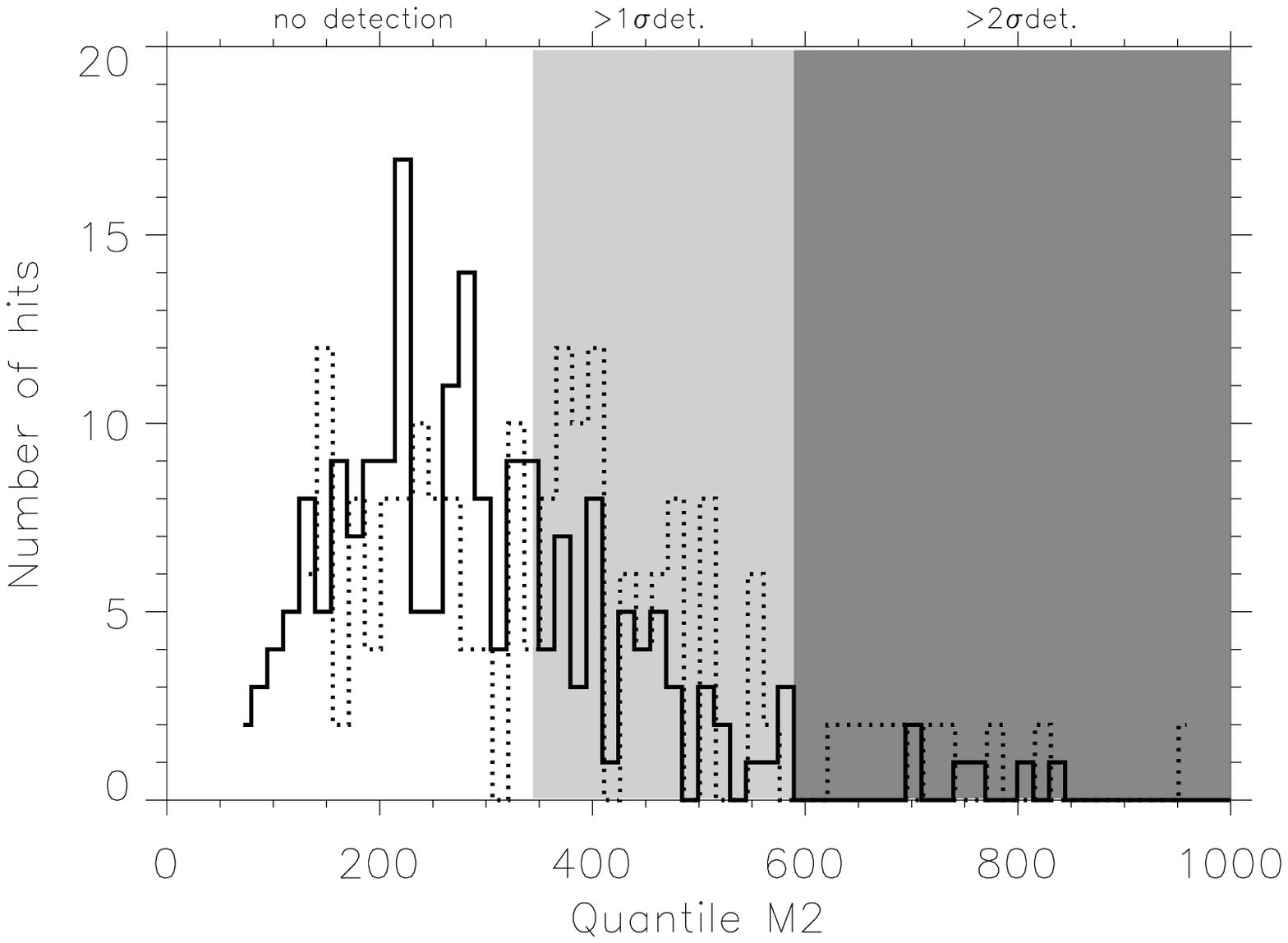,width=7cm,height=7cm}
\end{center}
\caption{Histogram with threshold levels for all three Minkowski functionals for Gaussian realizations (solid line) and non-Gaussian realizations with $%
f_{NL}=100$ (dotted line)(the $f_{NL}$ factor will be explained in section \ref{sect:comp}). All realizations have Planck-like noise and beam $%
20^{\prime}$. The shaded areas represent the $1\protect\sigma$ and $2\protect%
\sigma$ detection limits.}
\label{fig:hist_fnl300_mc}
\end{figure}

\begin{itemize}
\item Given an observed map, we estimate the power spectrum.

\item Using the estimated power spectrum, we generate 200 Gaussian
realizations \cite{healpix}

\item For each map we calculate the Minkowski functionals, using the tangent
planes as described above

\item We calibrate the quantiles (i.e. the threshold values at a given significance level) using the Monte Carlo simulations; the one and two sigma detection levels are shown in fig.\ref{fig:hist_fnl300_mc}.

\item We calculate our statistic (\ref{eq:quantile}) from the observed map and compare the result with the Monte-Carlo calibration; we are thus able to determine at which confidence level the map is Gaussian or not.
\end{itemize}

\begin{figure}[tbp]
\begin{center}
\leavevmode \includegraphics[width=7cm,height=7cm]{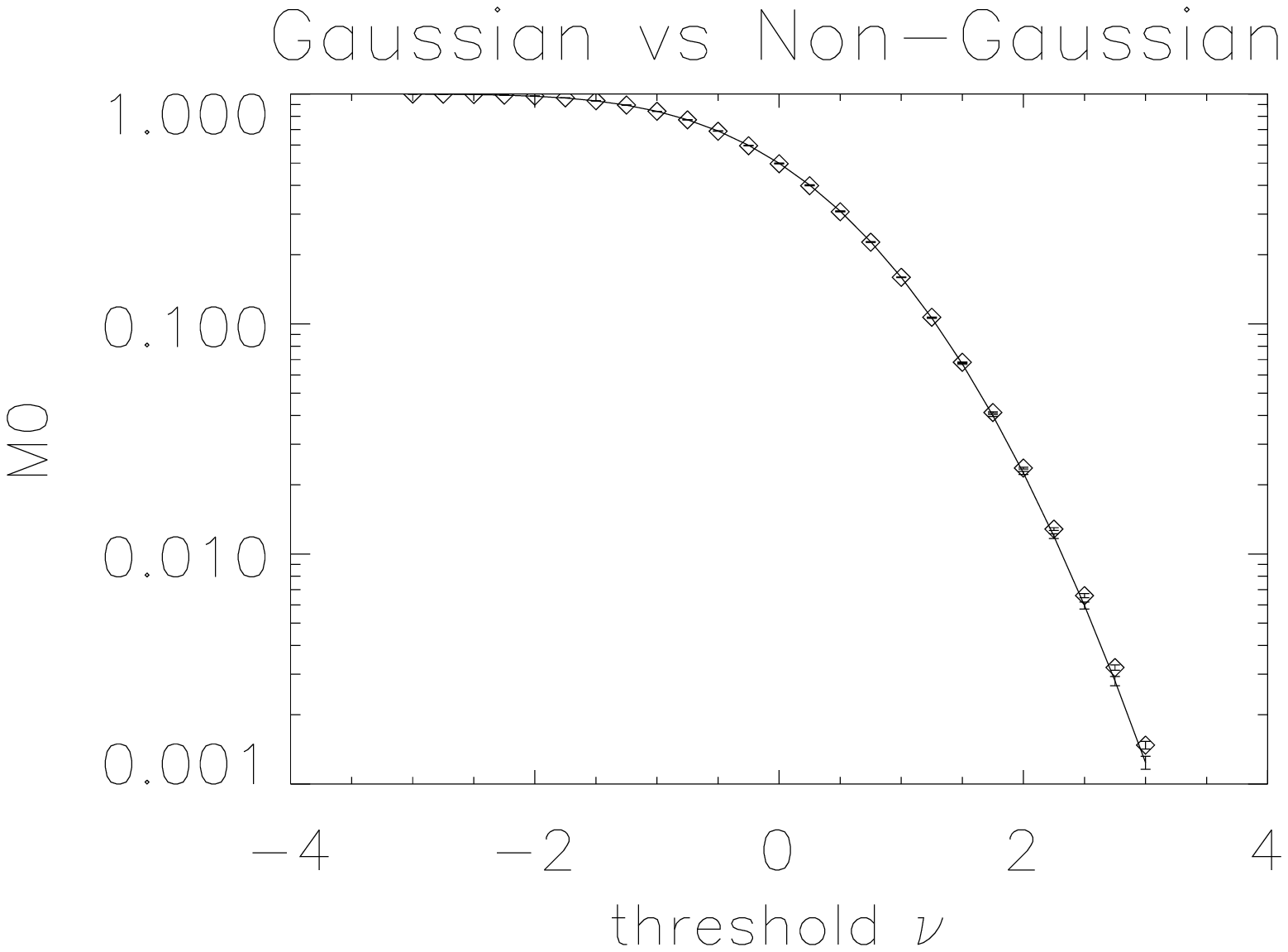} %
\includegraphics[width=7cm,height=7cm]{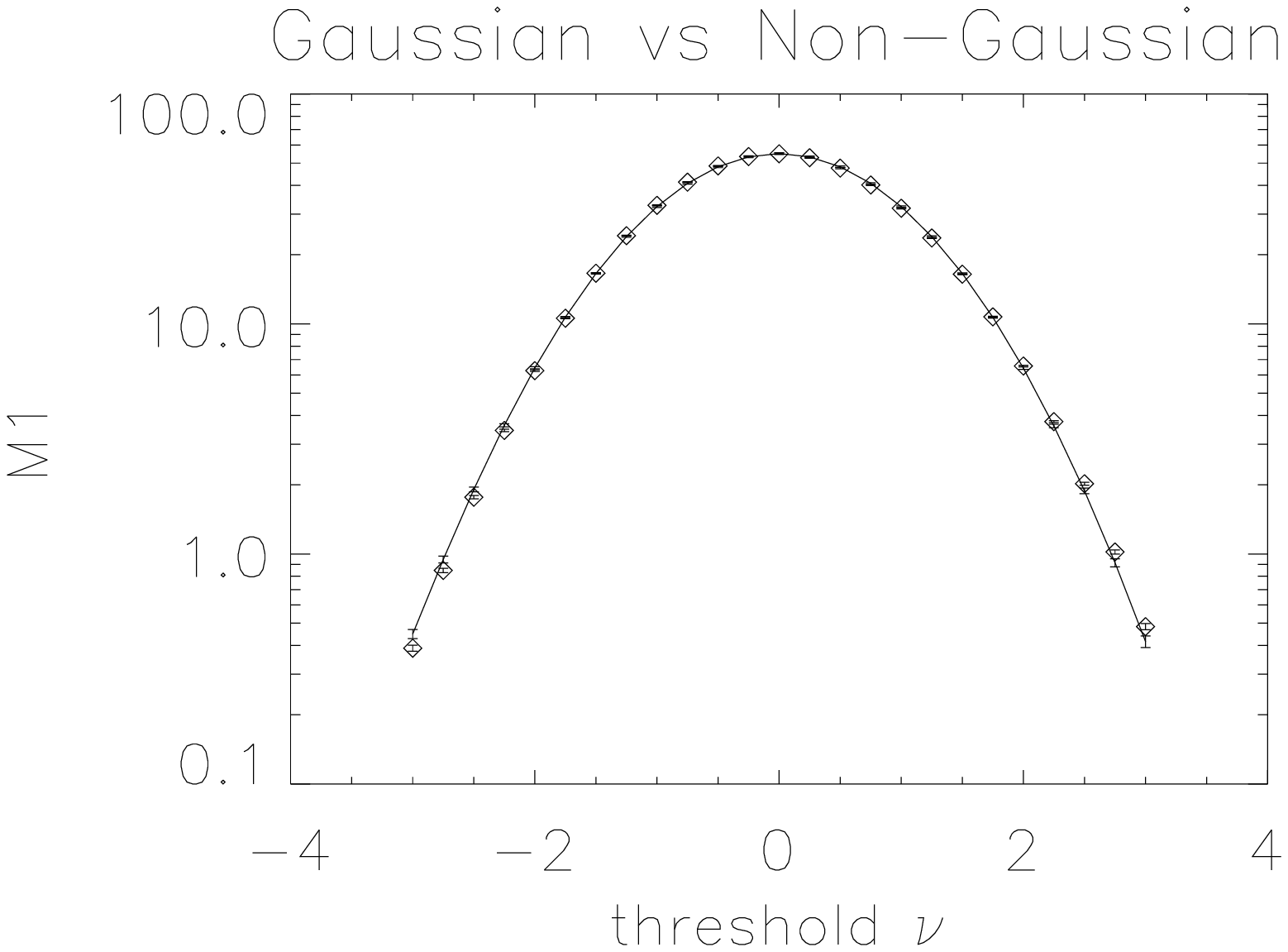} %
\includegraphics[width=7cm,height=7cm]{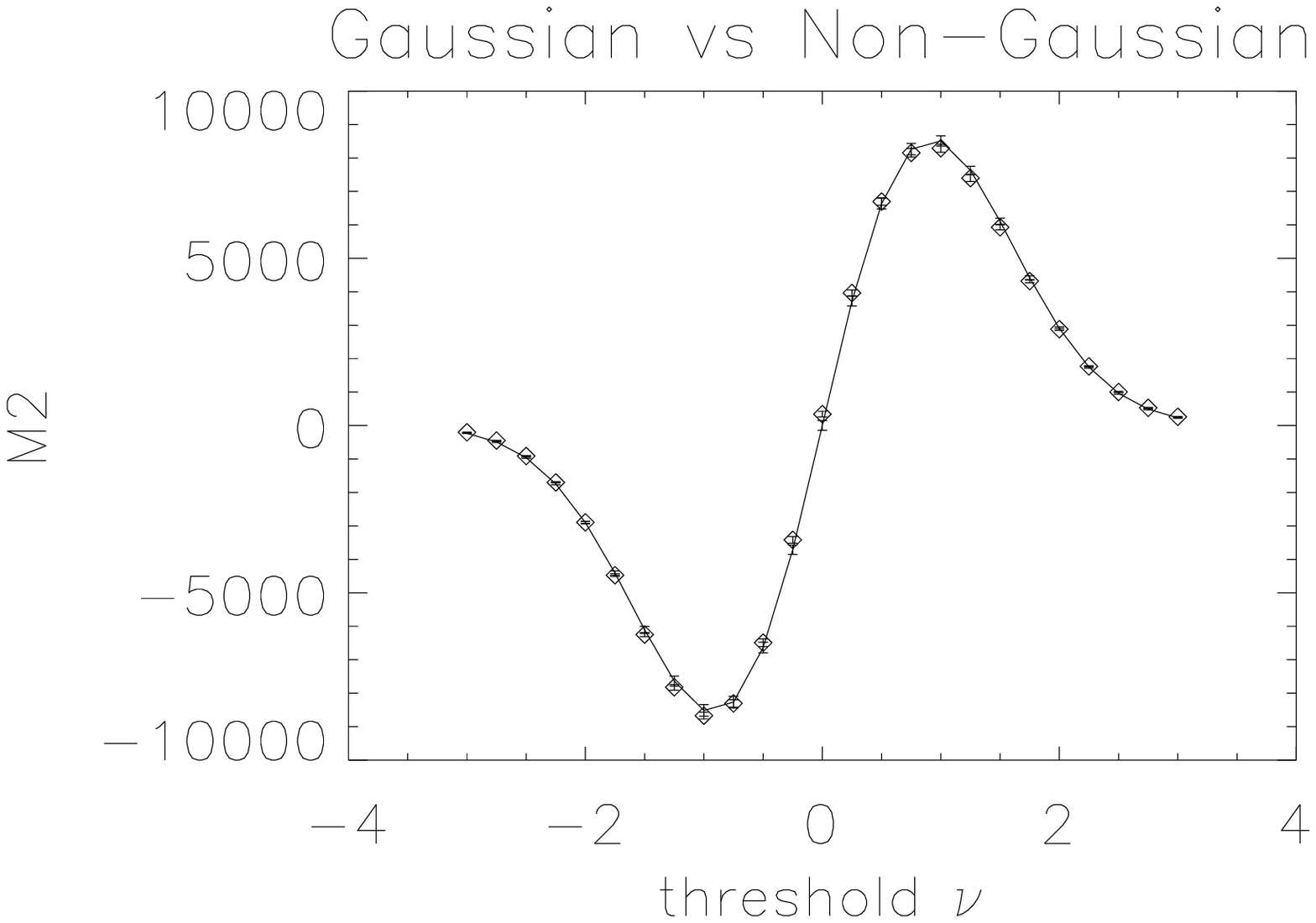}
\end{center}
\caption{Minkowski functionals averaged over 200 realization of Gaussian
maps (solid line) and 100 non-Gaussian maps (squares) with $f_{NL}=300$, Planck noise and beam 20'. The small $1\protect\sigma$ deviations are
also shown.}
\label{fig:Mink_fnl300_mc}
\end{figure}

\begin{figure}[tbp]
\begin{center}
\leavevmode \includegraphics[width=16cm,height=20cm]{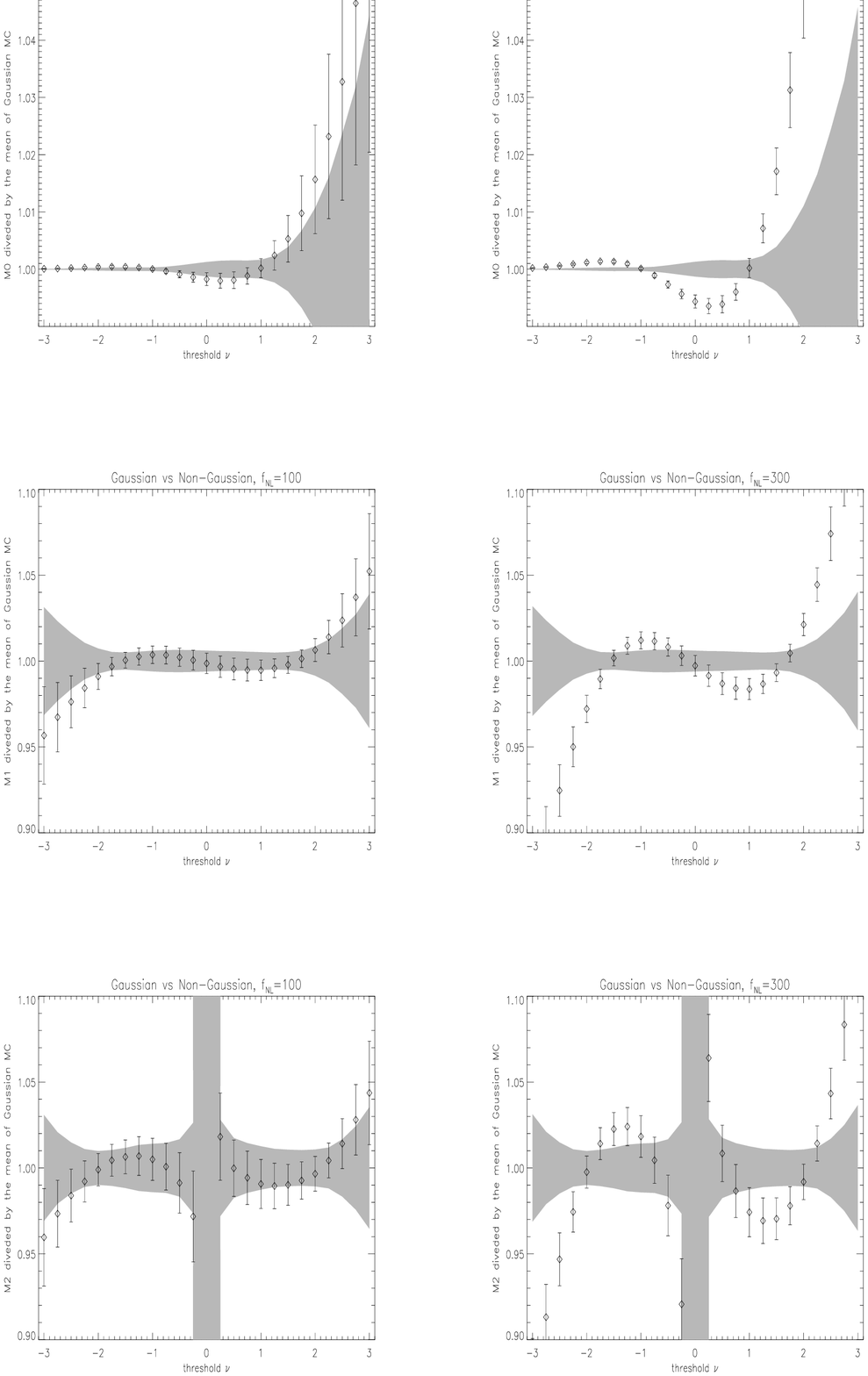}
\end{center}
\caption{Minkowski functionals (divided by the mean) averaged over 200 realization of Gaussian
maps (solid line) and 100 non-Gaussian maps (squares) with $f_{NL}=300/100$%
, Planck noise and beam 20'. The shaded bands show the $%
1\protect\sigma$ deviations of the Gaussian realizations and the error bars
show the $1\protect\sigma$ deviations for the non-Gaussian maps.}
\label{fig:merge}
\end{figure}

\section{The empirical process method}

\label{sect:ep} The details of the empirical process approach to detect
non-Gaussianity in the CMB were given in \cite{paper1,paper2,mar}. In short, the method consists of a family of tests which focus on the total distribution
of $a_{\ell m}$ and check for dependencies between $k$ $\ell$-rows. The
first step is to transform the spherical harmonic coefficients into
variables $u_{\ell m}$ which have an approximate uniform distribution
between $0$ and $1$, given that the $a_{\ell m}$ were initially Gaussian
distributed. This is done using the Smirnov transformation, defined as 
\[
u_{\ell 0}=\Phi _{1}\left(\frac{|a_{\ell 0}|^{2}}{\hat C_{\ell}}\right)\text{
, }u_{\ell m}=\Phi _{2}\left(\frac{2|a_{\ell m}|^{2}}{\hat C_{\ell}}\right)%
\text{ , }m=1,2,...,l,\text{ }l=1,2,...L\text{ ,} 
\]%
where $\Phi_n$ is the cumulative distribution function of a $\chi^2$ with $n$
degrees of freedom and $\hat C_\ell$ are the power spectrum coefficients
estimated from the data. The error introduced by using estimated $\hat
C_\ell $ instead of the real underlying $C_\ell$ is dealt with using a
bias-subtraction, as described in \cite{paper1}.\newline

Then the joint empirical distribution function for row $\ell$ is formed, 
\[
\widehat{F}_{\ell...\ell+\Delta_{\ell,k-1}}(\alpha _{1},...,\alpha _{k})=%
\frac{1}{(\ell+1)}\sum_{m=0}^{\ell}\left\{ \underline{\mathbf{1}}(\widehat{u}%
_{\ell m}\leq \alpha _{1})\prod_{i=2}^{k}\underline{\mathbf{1}}(\widehat{u}%
_{\ell+\Delta_{\ell,i-1},m+\Delta _{mi}}\leq \alpha _{i})\right\} \text{ , }%
\Delta _{mi}\geq 0\text{ ,} 
\]
where $\Delta_{\ell,i}$ determines the spacing between the rows for which
the dependencies are tested and $\Delta_{mi}$ denotes the difference in $m$
for row $i$. The parameters $\alpha_i$ run over the interval $[0,1]$. The
empirical process is expressed using the centered and rescaled $\widehat{F}%
_{\ell...\ell+\Delta_{\ell,k-1}}$ given as 
\[
\widehat{G}_{\ell...\ell+\Delta_{\ell,k-1}}(\alpha _{1},...,\alpha _{k})=%
\sqrt{(\ell+1)}\left\{ \widehat{F}_{\ell...\ell+\Delta_{\ell,k-1}}(\alpha
_{1},...,\alpha _{k})-\prod_{i=1}^{k}\alpha _{i}\right\} \text{ .} 
\]
The intuition behind this procedure is as follows: if the $a_{\ell m}$s are
Gaussian, $\widehat{G}$ converges to a well-defined limiting process, whose
distribution can be readily tabulated. On the other hand, for non-Gaussian $%
a_{\ell m}$s $\{\widehat{F}(\alpha_{1},...,\alpha_{k})-\prod_{i=1}^{k}%
\alpha_{i}\}$ and thereby $\widehat{G}$ will take `high' values over some
parts of $\alpha$-space. Thus, the analysis of some appropriate functional
of $\widehat{G}$ can be used to detect non-Gaussianity. To combine the
information over all multipoles into one statistic, we define 
\begin{equation}
\label{eq:kla}
\widehat{K}_{L}(\alpha _{1},...,\alpha _{k},r)=\frac{1}{\sqrt{L-\Delta
_{\ell,k-1}}}\sum_{\ell=1}^{[(L-\Delta _{k-1})r]}\widehat{G}%
_{\ell,...,\ell+\Delta _{\ell,k-1}}(\alpha _{1},...,\alpha _{k})\text{ ,} 
\end{equation}
where $L$ is the highest multipole where the data is signal dominated.

The method can then be summarized as follows: the distribution of $\mathrm{%
sup}|\widehat{K}_{L}|$ is found using Monte-Carlo simulations of Gaussian
distributed $a_{\ell m}$. Then, for a given observed set of $a_{\ell m}$,
the value $k_\mathrm{max}=\mathrm{sup}|\widehat{K}_{L}|$ is found and
compared to the distribution obtained from Monte-Carlo. The consistency of
the data with a Gaussian distribution can then be estimated to any suitable $%
\sigma$-level. In \cite{paper2} this simple approach was extended in three
different ways. First of all, the fact that the above explained estimator is
not rotationally invariant is exploited, using the $k_\mathrm{max}$ value
averaged over many rotations. Each rotation can be viewed as a resampling of
the $a_{\ell m}$. Secondly, we introduced three variations of the test,
taking into account, not only the modulus of the $a_{\ell m}$ but also the
phases. Finally, experimental effects like noise and galactic cut was
accounted for using Monte Carlo calibration of the $k_\mathrm{max}$
distribution with these effects included. It should be noted that the rotated maps are clearly dependent and the resulting statistic may thus depend on the shape of the angular power spectrum.

\section{Test of non-Gaussianity in wavelet space}
\label{sect:wav}

A third space where one could look for non-Gaussianity is the wavelet space. The use of wavelets for non-Gaussianity tests of the CMB has been investigated by several authors \cite{barreiro1,barreiro3,mg,cobewav} and turns out to be a very powerful tool. We will here just briefly describe the wavelet method, and refer to the above references for more details.

An isotropic wavelet can be defined as
\begin{equation}
\Psi(\vec{x};\vec{b},R)=\frac{1}{R}\psi\left(\frac{|\vec{x}-\vec{b}|}{R}\right)
\end{equation}
with the properties
\begin{equation}
\int d\vec{x}\:\psi(x) = 0
\end{equation}
\begin{equation}
\int d\omega\: \frac{\psi^2(\omega)}{\omega}\:\:<\infty 
\end{equation}

where $x=|\vec{x}|$, R represents a scale and b a translation. The Fourier transform of the wavelet is represented by $\psi(\omega)$. We will focus on the mexican hat wavelets given by:

\begin{equation}
\Psi(\vec{x};\vec{b},R)=\frac{1}{(2\pi)^{1/2}R}\left[2-(\frac{|\vec{x}-\vec{b}|}{R})^2\right]e^{-|\vec{x}-\vec{b}|^2/2R^2}
\end{equation} 
From the wavelet transform of a function $f(\vec{x})$ one can obtain the wavelet coefficients :
\begin{equation}
w(\vec{b},R)=\int d\vec{x}\psi(\vec{x};\vec{b},R)f(\vec{x})
\end{equation} 
and if $f(\vec{x})$ is Gaussian, $w(\vec{b},R)$ will be Gaussian as well. We will here use the CMB temperature fluctuation field as the $f(\vec{x})$ function. We will implement the non-Gaussianity test in wavelet space as we did for the Minkowski functionals:
\begin{itemize}
\item we generate a set of Gaussian CMB maps for calibration and a set of non-Gaussian maps for testing  
\item we cut tangent planes $(12^\circ \times 12^\circ)$ 
\item we calculate for each plane the coefficients $w(\vec{b},R)$
\item we evaluate the skewness of the wavelet coefficients for each sky.
\item finally, using the skewness of the wavelets from the Gaussian maps, we define the one and two $\sigma$ detection levels as described above for the other methods.
\end{itemize}

\section{Comparison and combined test}
\label{sect:comp}

This section aims at comparing the different methods described above.
Applying the methods on the same maps , we will first compare the number of detections. We will in this paper use the non-standard inflationary model described in \cite{sabino,michele} which has the non-linear coupling parameter $f_{NL}$ as a measure of the strength of non-Gaussianity.  We generated 100 `observed' skies with Planck-like noise (LFI 100 GHz), beam 20', pixel-size $\simeq$ 6' (Nside 512 in Healpix language), using a pure Sachs-Wolfe spectrum with $f_{NL}$ values of 300 and 100. In figures \ref{fig:Mink_fnl1000_mc}, \ref{fig:Mink_fnl300_mc} and \ref{fig:merge} one can see the behavior of the Minkowski functionals in the presence of a non-zero $f_{NL}$. In tables \ref{tab:M0} and \ref{tab:fnl100} we show the rejection rates. As the first Minkowski
functional was giving the best results in this tests, we will
focus only on $M_0$ for this kind of non-Gaussianity. In table \ref{tab:wav} we list the results of the wavelet test on the same maps (for the wavelets we used the parameter $R=22.5'$). For individual results of the empirical process test, we refer to \cite{paper2}.\newline

\begin{table}[h!]
\caption{The quantile levels which determine the one and two $\sigma$ detections and the rejection rates for non-standard inflationary models with $f_{NL}=300$}
\label{tab:M0}
\begin{center}
\begin{tabular}{|c||c|c||c||c|c||c||c|c|}
\hline
\multicolumn{9}{|c|}{{\ quantile detection limits}} \\ \hline\hline
$I_0 $ & $1\sigma$ & 3.6 $10^{-3}$ & $I_1$ & $1\sigma$ & 0.286 &$I_2$  & $1\sigma$ & 33.2 \\ 
 & $2\sigma$ & 6.3 $10^{-3}$ &  & $2\sigma$ & 0.524 &  & $%
2\sigma$ & 49.7 \\  \hline\hline
\multicolumn{9}{|c|}{ Rejection rates $\%$} \\ \hline\hline
$I_0$ & $1\sigma$ & 100\% & $I_1$ & $1\sigma$ & 100\% &$I_2$  & $1\sigma$ & 100\% \\  & $2\sigma$ & 100\% &  & $2\sigma$ & 71\% &  & $2\sigma$ & 84\%
\\  \hline\hline
\end{tabular}%
\end{center}
\end{table}

\begin{table}[h!]
\caption{The quantile levels which determine the one and two $\sigma$ detections and the rejection rates for non-standard inflationary models with $f_{NL}=100$}
\label{tab:fnl100}
\begin{center}
\begin{tabular}{|c||c|c||c||c|c||c||c|c|}
\hline
\multicolumn{9}{|c|}{{\ quantile detection limits}} \\ \hline\hline
$I_0$ & $1\sigma$ & 3.6 $10^{-3}$ &$I_1$ & $1\sigma$ & 0.286 & $I_2$ & $1\sigma$ & 33.2 \\ & $2\sigma$ & 6.3 $10^{-3}$ &  & $2\sigma$ & 0.524 &  & $%
2\sigma$ & 49.7 \\  \hline\hline
\multicolumn{9}{|c|}{Rejection rates $\%$} \\ \hline\hline
$I_0$ & $1\sigma$ & 68\% & $I_1$ & $1\sigma$ & 52\% & $I_2$  & $1\sigma$ & 57\% \\  & $2\sigma$ & 35\% &  & $2\sigma$ & 8\% &  & $2\sigma$ & 
8\% \\  \hline\hline
\end{tabular}%
\end{center}
\end{table}

\begin{table}
\caption{wavelets test: the rejection rates at one and two $\sigma$ for non-standard inflationary models with $f_{NL}=100$}
\label{tab:wav}

\begin{center}
\begin{tabular}{|l|l|}

\hline
\multicolumn{2} {|l|} {$f_{NL}=$  100}  \\ 
\hline
Confidence Level & Rejection Rate\\
\hline
1 $\sigma $ & 89\% \\
2 $\sigma $ & 57\% \\ 
\hline
\end{tabular}
\end{center}
\end{table}

Table \ref{tab:totres} shows the number of detections at the different
levels, using $M_0$ and the empirical process method on maps with $%
f_{NL}=100 $ and $f_{NL}=300$. The power of the two procedures appears very close. However, analyzing the individual maps, we find
that only one third of the maps detected at $2\sigma$ are the same for the
two tests. This leads to the idea of implementing a combined test.\newline

\begin{center}
\begin{table}[tbp]
\caption{COMPARISON AND COMBINED TEST}
\label{tab:totres}%
\begin{ruledtabular}
\begin{tabular}{|l|l|l|l|l|l|}
TEST & emp. proc & & $M_0$ & & comb \\
\hline
$f_{NL}$ & 100 &  300 &  100 & 300 & 100  \\ 
\hline
1 $\sigma $ & 50\% & 95\% & 68\% & 100\% & 74\% \\
2 $\sigma $ & 29\% & 87\% & 35\% & 100\% & 35\% \\ 
\end{tabular}
\end{ruledtabular}
\end{table}
\end{center}

\begin{figure}[tbp]
\leavevmode
\par
\begin{center}
\leavevmode \epsfig {file=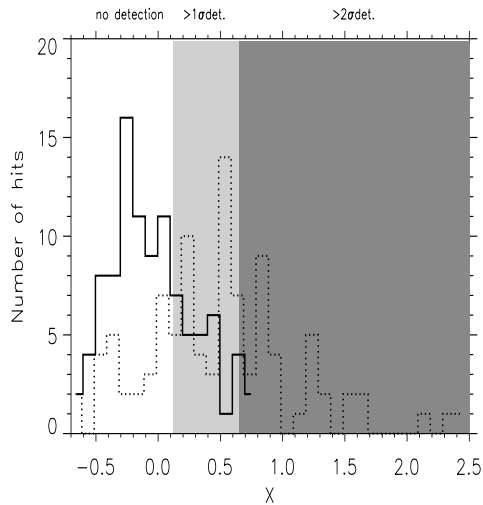,width=7cm,height=7cm}
\end{center}
\caption{Histogram of the combined estimator $x$ (empirical process and Minkowksi functionals) for Gaussian realizations
(solid line) and non-Gaussian realizations with $f_{NL}=100$ (dotted line).
All realizations have Planck-like noise and beam $20^{\prime}$. The shaded
areas represent the $1\protect\sigma$ and $2\protect\sigma$ detection limits.}
\label{fig:hist_comb_mc}
\end{figure}

For the combined test we suggest to use an indicator consisting of $I_0$
from the Minkowski functionals and $k_\textrm{max}$ from the empirical
process. We chose to normalize the $I_0$ and $k_\textrm{max}$ so that they
both have mean zero and variance one, using Monte-Carlo simulations of
Gaussian maps. In this way, the two values can be averaged: 
\begin{equation}
\label{eq:comb}
x=w_1\tilde I_0+w_2\tilde k_\textrm{max},
\end{equation}
where 
\begin{equation}
\tilde I_0=(I_0-\langle I_0\rangle)/\sqrt{\langle I_0^2\rangle-\langle
I_0\rangle^2},
\end{equation}
and 
\begin{equation}
\tilde k_\textrm{max}=(k_\textrm{max}-\langle k_\textrm{max}\rangle)/\sqrt{%
\langle k_\textrm{max}^2\rangle-\langle k_\textrm{max}\rangle^2}.
\end{equation}
Here $\langle \rangle$ means mean value taken over 100 Gaussian simulations.
The weights $w_1$ and $w_2$ were chosen proportionally to the power of each procedure. Using the rejection rates for $f_{NL}=100$ in table \ref{tab:fnl100}, we arrived at $w_1\approx0.6$ and $w_2\approx0.4$. Of course, the threshold value for $x$ needs to be evaluated anew. In figure (\ref{fig:hist_comb_mc}) we plot the distribution of $x$ for the Gaussian and $%
f_{NL}=100$ non-Gaussian maps.\newline

By a similar motivation, it is natural to combine also the wavelet method into a single procedure; it is easy to see (tables \ref{tab:fnl100} and \ref{tab:wav}) that the detection rate at $2\sigma$ is about two times higher for the wavelets, warranting the wavelet coefficient a very high weight in the combined analysis. Inspecting table \ref{tab:totres}, we detect a moderate improvement in the detection rate when combining Minkowski functionals and empirical process. The result of the empirical process + Minkowski functionals + wavelets combined test is shown in table \ref{tab:allcomb}. A significant improvement of the number of detections at both confidence levels is evident; note that the combined procedure is to some extent model dependent, as the weights we used were tabulated from specific non-Gaussian models.

\begin{table}
\caption{The results of the combined test using all three methods: Empirical process, Minkowski functionals and wavelets for $f_{NL}=100$}
\label{tab:allcomb}

\begin{center}
\begin{tabular}{|l|l|}

\hline
\multicolumn{2} {|l|} {$f_{NL}=$  100}  \\ 
\hline
Confidence Level & Rejection Rate\\
\hline
1 $\sigma $ & 100\% \\
2 $\sigma $ & 78\% \\ 
\hline
\end{tabular}
\end{center}
\end{table}

\section{Point sources}
\label{sect:point}

\begin{figure}[tbp]
\leavevmode
\begin{center}

\epsfig {file= 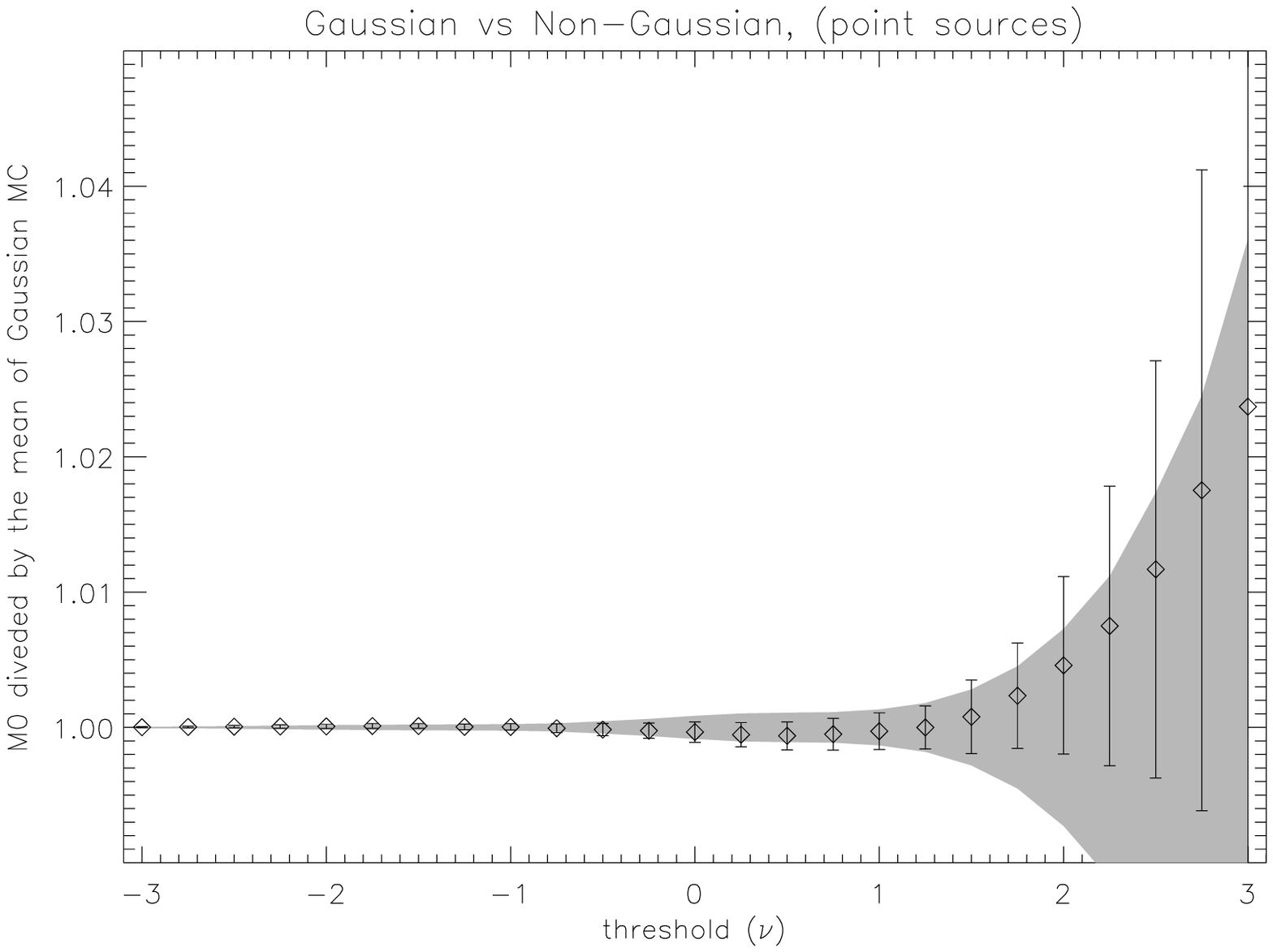,width=7cm,height=7cm}
\epsfig {file= 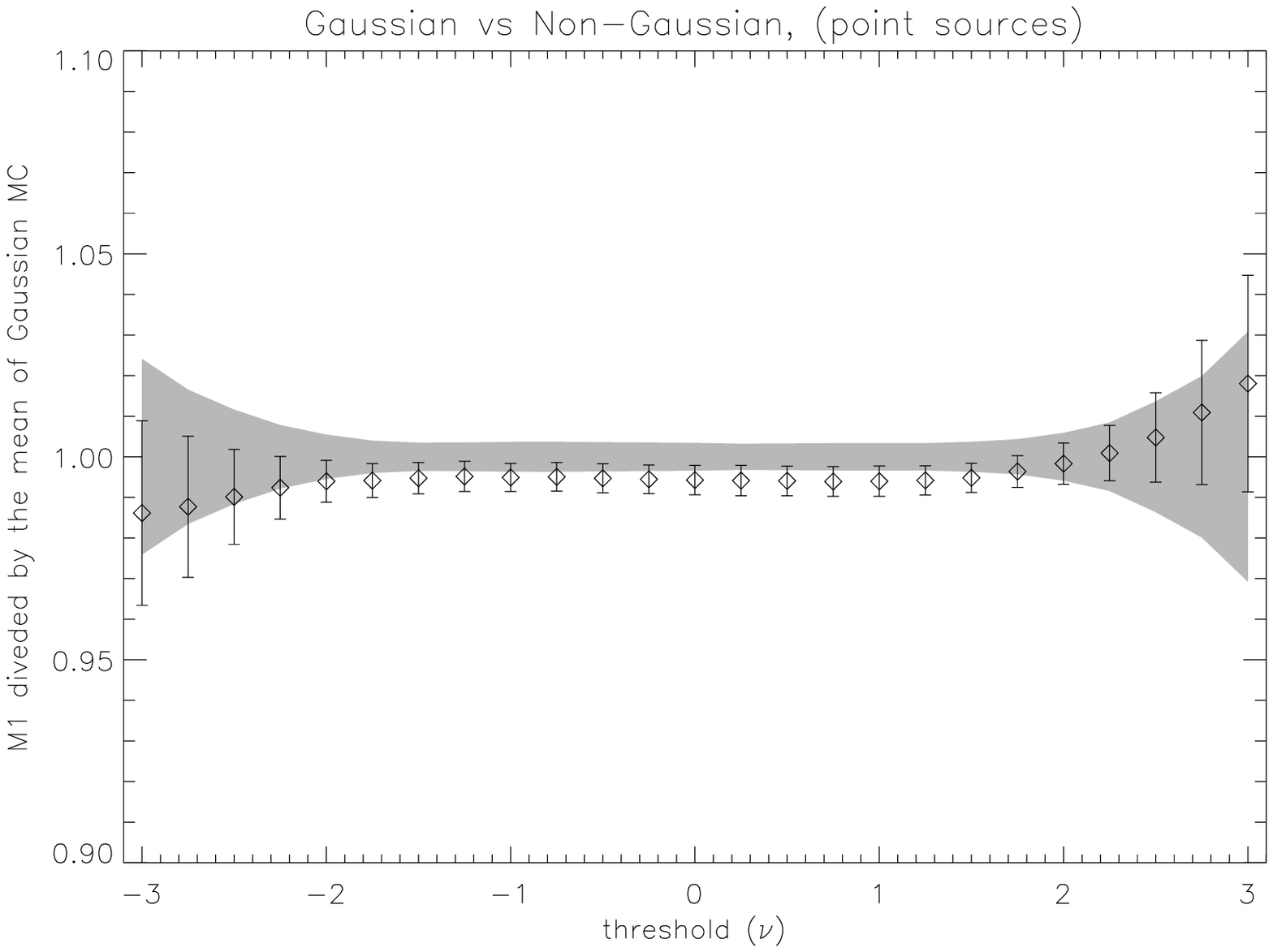,width=7cm,height=7cm}
\epsfig {file= 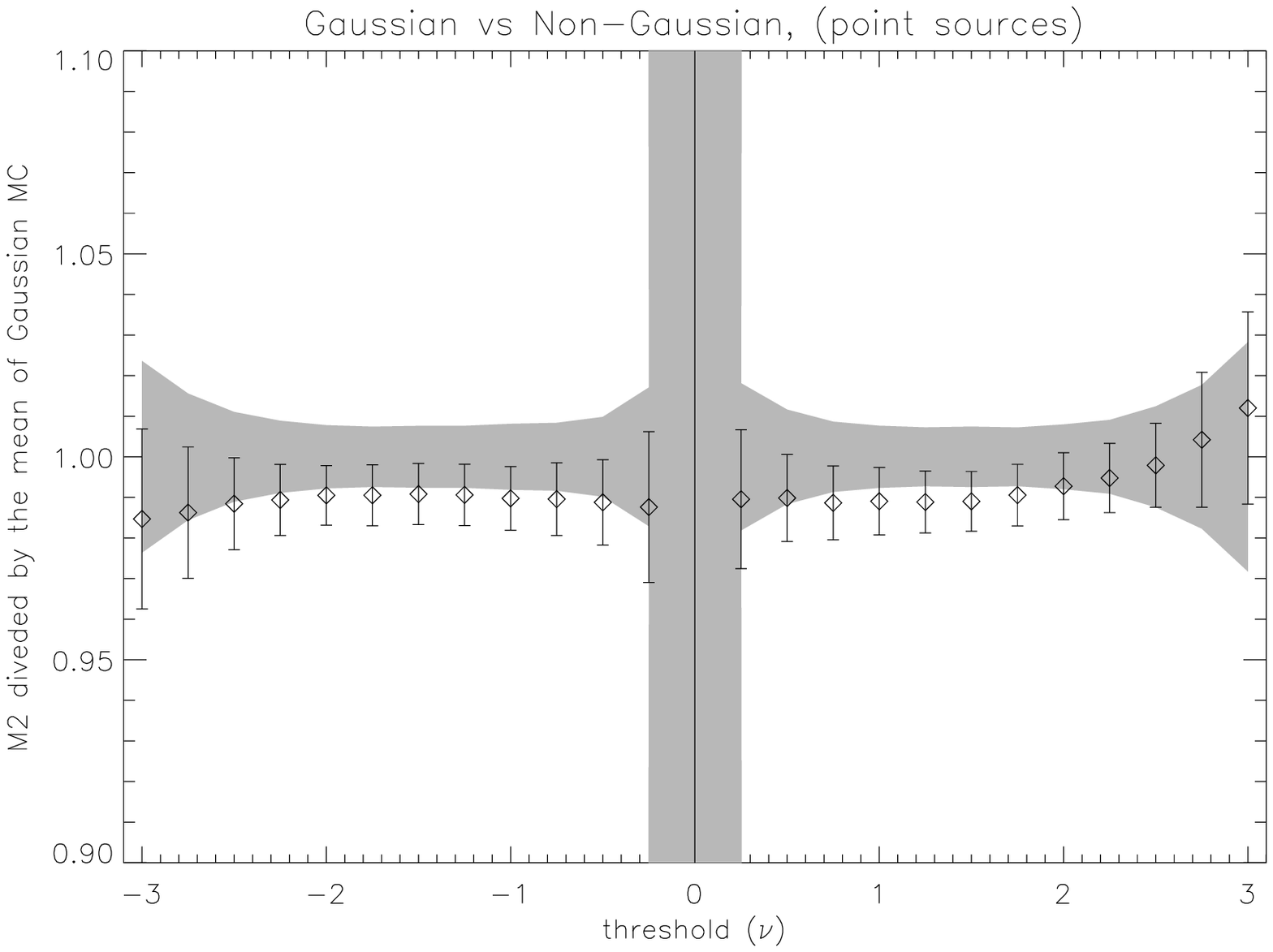,width=7cm,height=7cm}

\caption{Minkowski functionals (divided by the mean) averaged over 200 realization of Gaussian maps (solid line) and 100 non-Gaussian maps (squares) with point sources, Planck noise and beam 20'. The shaded bands show the $1\sigma$ deviations of the Gaussian realizations and the error bars show the $1\sigma$ deviations for the non-Gaussian maps.}
\label{fig:sorg}
\end{center}
\end{figure}

We also simulated maps with point sources (with noise and beam as given above) to compare the power of the methods on a different kind of non-Gaussianity. We generated a toy model of point sources with a distribution building on formula (1) in \cite{sorgenti} and formulae (1) and (2) in \cite{tegmark}. In table \ref{tab:sorg} we show the results for the Minkowski functionals. We see immediately that for this kind of non-Gaussianity, the first Minkowski functional is not sensitive, whereas the other two functionals show a good rejection rate. This suggests that we might be able to discriminate between these two types of non-Gaussianity. The first Minkowski functional can be used to trace primordial non-Gaussianity with little influence from the point sources. On the other hand, the presence of point sources will show up in the second and third Minkowski functionals, which are only weakly influenced by primordial non-Gaussianity. In Figure (\ref{fig:sorg}) we show the shape of the Minkowski functionals in the presence of point sources. Note that the deviations from the Gaussian mean are different than in the case of primordial non-Gaussianity (Figure \ref{fig:Mink_fnl300_mc}). The point sources manifest themselves mainly as an offset in $M_1$ and $M_2$, consistent with what was observed for weak lensing \cite{Schmalzing2}. However as seen in figure (\ref{fig:merge}) for non-standard inflation the curve has a particular shape. \\

One could imagine combining $M_1$ and $M_2$ in order to strengthen the power of the test, similarly to what we have done above. However, it turns out that the maps detected by $M_2$ are contained within the maps detected by $M_1$, so that there is no additional information in combining the two estimators.

\begin{table}[h!]
\caption{The quantile levels which determine the one and two $\sigma$ detections and the rejection rates for maps contaminated by point sources}
\label{tab:sorg}
\begin{center}
\begin{tabular}{|c||c|c||c||c|c||c||c|c|}

\hline
\multicolumn{9}{|c|}{{\ quantile detection limits}}\\
\hline\hline
$I_0 $    &$1\sigma$&  1.1   $10^{-3}$  &  $I_1$      &$1\sigma$&  0.415 &      $I_2$  &$1\sigma$& 207.6\\
 &$2\sigma$& 1.7 $10^{-3}$  & &$2\sigma$&  0.770 &  &$2\sigma$& 328.3\\
\hline\hline

 \multicolumn{9}{|c|}{Rejection rates $\%$} \\
\hline\hline
  $I_0$    &$1\sigma$&  36\%  &	    $I_1$    &$1\sigma$&  81\%     & $I_2$     &$1\sigma$&  75\%\\  &$2\sigma$&  8\% &   &$2\sigma$&  45\% &  &$2\sigma$&  33\%\\

\hline

\end{tabular}
\end{center}
\end{table}

 For the empirical process method, there was no detection for any of the tests, univariate, bivariate or trivariate. The numbers $k_{max}$ obtained for the maps with point sources were consistent with those for Gaussian maps. Also for the wavelet test, the number of detections was very small. Note that wavelets can be useful for detection of bright point sources \cite{points}, but in our source model these were excluded. This might suggest that the empirical process methods and the skewness of the wavelets can be used to probe primordial non-Gaussianity without confusion from point sources, making the combined test presented above more robust.\\

\section{Comments and conclusions}
\label{sect:concl}

We have compared three methods for detecting non-Gaussianity in observations of the cosmic microwave background. The methods were applied to non-Gaussian maps with two different kinds of non-Gaussianity, primordial non-Gaussianity with a varying $f_{NL}$ and point sources. It is important to note that the non-Gaussian maps used in this article were generated taking into account only the Sachs-Wolfe effect. In future work we will study maps where the full radiative transfer equations have been applied. For the time being, we stress that our results are broadly consistent with the power of the procedures adopted for WMAP data analysis. More precisely, a detailed comparison is unfeasible, as we are assuming a simplified non-Gaussian model (no radiative transfer) and Planck LFI like noise and beam. Moreover we are restricting the analysis to the first 500 multipoles . Broadly speaking, however,  Monte-Carlo simulations suggest that a value of $f_{NL}$ about 100 represents the lower limit that can be detected at a $2\sigma$ level, by using a combined procedure: this seems consistent with the value $f_{NL}=139$ reported in (\cite{WMAPKomatsu}). As an estimator of non-Gaussianity for the Minkowski functionals we have introduced a statistic which gathers the information from different thresholds (eq.\ref{eq:quantile}). The estimator for the empirical process method is $k_{max}$, the maximum value of the function $K(\alpha,r)$ obtained from a given map (eq.\ref{eq:kla}) using the trivariate test.  For the test in wavelet space, we use the skewness of wavelet coefficients.\\

For the primordial non-Gaussianity, $M_0$ and the empirical process method have a similar rejection rate, whereas $M_1$ and $M_2$  showed less power. On the other hand, for the maps with point sources, $M_1$ and $M_2$ gave the best results, whereas $M_0$ and the empirical process method had no rejections. The fact that the first Minkowski functional shows little power in the presence of point sources is hardly surprising. Indeed this statistic depends only on the pixel by pixel temperature values and hence is not at all affected by discontinuities in the map. The converse is clearly true for the other Minkowski functionals, which are sensitive to the local morphology of the maps. For the empirical process, we simply note that spikes in real space are erased in harmonic space. It is also important to stress that our results depend heavily upon the nature of non-Gaussianity; in particular some preliminary exploration of Monte-Carlo evidence from non-physical toy models suggests that the power of these procedures need not be close, in general. This strengthens the case for (weighted) multiple/combined procedures; the combined procedures seem to show a marked improvement in the power of the test. However, the pixel-, harmonic- and wavelet-space methods, despite carrying complementary statistical information, should not be viewed neither as orthogonal nor as independent, so that some care is needed when merging them into a single statistic. In any case, the fact that different methods detect different kinds of  non-Gaussianity can be viewed as an advantage, in the sense that, for instance, primordial non-Gaussianity can be detected without confusion from point sources.

\begin{acknowledgments}

We are thankful to Michele Liguori and Sabino Matarrese for supplying maps with primordial non-Gaussianity. We also wish to thank Paolo Natoli for suggestions and useful discussions. This research used resources of the National Energy Research Scientific Computing Center, which is supported by the Office of Science of the U.S. Department of Energy under Contract No. DE-AC03-76SF00098. We acknowledge the use of the Healpix package \cite{healpix}.  

\end{acknowledgments}



\begin{thebibliography}{44}
\expandafter\ifx\csname natexlab\endcsname\relax\def\natexlab#1{#1}\fi
\expandafter\ifx\csname bibnamefont\endcsname\relax
  \def\bibnamefont#1{#1}\fi
\expandafter\ifx\csname bibfnamefont\endcsname\relax
  \def\bibfnamefont#1{#1}\fi
\expandafter\ifx\csname citenamefont\endcsname\relax
  \def\citenamefont#1{#1}\fi
\expandafter\ifx\csname url\endcsname\relax
  \def\url#1{\texttt{#1}}\fi
\expandafter\ifx\csname urlprefix\endcsname\relax\def\urlprefix{URL }\fi
\providecommand{\bibinfo}[2]{#2}
\providecommand{\eprint}[2][]{\url{#2}}

\bibitem[{\citenamefont{Martin et~al.}(2000)\citenamefont{Martin, Riazuelo, and
  Sakellariadou}}]{nongi1}
\bibinfo{author}{\bibfnamefont{J.}~\bibnamefont{Martin}},
  \bibinfo{author}{\bibfnamefont{A.}~\bibnamefont{Riazuelo}}, \bibnamefont{and}
  \bibinfo{author}{\bibfnamefont{M.}~\bibnamefont{Sakellariadou}},
  \bibinfo{journal}{Phys. Rev.} \textbf{\bibinfo{volume}{D61}},
  \bibinfo{pages}{083518} (\bibinfo{year}{2000}).

\bibitem[{\citenamefont{Contaldi et~al.}(2000)\citenamefont{Contaldi, Bean, and
  Magueijo}}]{nongi2}
\bibinfo{author}{\bibfnamefont{C.~R.} \bibnamefont{Contaldi}},
  \bibinfo{author}{\bibfnamefont{R.}~\bibnamefont{Bean}}, \bibnamefont{and}
  \bibinfo{author}{\bibfnamefont{J.}~\bibnamefont{Magueijo}},
  \bibinfo{journal}{Phys. Lett.} \textbf{\bibinfo{volume}{B468}},
  \bibinfo{pages}{189} (\bibinfo{year}{2000}).

\bibitem[{\citenamefont{Linde and Mukhanov}(1997)}]{nongi3}
\bibinfo{author}{\bibfnamefont{A.}~\bibnamefont{Linde}} \bibnamefont{and}
  \bibinfo{author}{\bibfnamefont{V.}~\bibnamefont{Mukhanov}},
  \bibinfo{journal}{Phys. Rev.} \textbf{\bibinfo{volume}{D56}},
  \bibinfo{pages}{535} (\bibinfo{year}{1997}).

\bibitem[{\citenamefont{Gupta et~al.}(2002)\citenamefont{Gupta, Berera,
  Heavens, and Matarrese}}]{nongi5}
\bibinfo{author}{\bibfnamefont{S.}~\bibnamefont{Gupta}},
  \bibinfo{author}{\bibfnamefont{A.}~\bibnamefont{Berera}},
  \bibinfo{author}{\bibfnamefont{A.~F.} \bibnamefont{Heavens}},
  \bibnamefont{and}
  \bibinfo{author}{\bibfnamefont{S.}~\bibnamefont{Matarrese}},
  \bibinfo{journal}{Phys. Rev.} \textbf{\bibinfo{volume}{D66}},
  \bibinfo{pages}{043510} (\bibinfo{year}{2002}).

\bibitem[{\citenamefont{Gangui et~al.}(2002)\citenamefont{Gangui, Martin, and
  Sakellariadou}}]{nongi6}
\bibinfo{author}{\bibfnamefont{A.}~\bibnamefont{Gangui}},
  \bibinfo{author}{\bibfnamefont{J.}~\bibnamefont{Martin}}, \bibnamefont{and}
  \bibinfo{author}{\bibfnamefont{M.}~\bibnamefont{Sakellariadou}},
  \bibinfo{journal}{Phys. Rev.} \textbf{\bibinfo{volume}{D66}},
  \bibinfo{pages}{083502} (\bibinfo{year}{2002}).

\bibitem[{\citenamefont{Bartolo et~al.}(2002)\citenamefont{Bartolo, Matarrese,
  and Riotto}}]{sabino}
\bibinfo{author}{\bibfnamefont{N.}~\bibnamefont{Bartolo}},
  \bibinfo{author}{\bibfnamefont{S.}~\bibnamefont{Matarrese}},
  \bibnamefont{and} \bibinfo{author}{\bibfnamefont{A.}~\bibnamefont{Riotto}},
  \bibinfo{journal}{Phys. Rev.} \textbf{\bibinfo{volume}{D65}},
  \bibinfo{pages}{103505} (\bibinfo{year}{2002}).

\bibitem[{\citenamefont{Liguori et~al.}(2003)\citenamefont{Liguori, Matarrese, and
  Moscardini}}]{michele}
\bibinfo{author}{\bibfnamefont{M.}~\bibnamefont{Liguori}},
  \bibinfo{author}{\bibfnamefont{S.}~\bibnamefont{Matarrese}},
  \bibnamefont{and}
  \bibinfo{author}{\bibfnamefont{L.}~\bibnamefont{Moscardini}},
  \bibinfo{journal}{ApJ} \textbf{\bibinfo{volume}{597}},
  \bibinfo{pages}{57} (\bibinfo{year}{2003}).

\bibitem[{\citenamefont{Gangui et~al.}(2001)\citenamefont{Gangui, Pogosian, and
  Winitzki}}]{strings}
\bibinfo{author}{\bibfnamefont{A.}~\bibnamefont{Gangui}},
  \bibinfo{author}{\bibfnamefont{L.}~\bibnamefont{Pogosian}}, \bibnamefont{and}
  \bibinfo{author}{\bibfnamefont{S.}~\bibnamefont{Winitzki}},
  \bibinfo{journal}{Phys. Rev.} \textbf{\bibinfo{volume}{D64}},
  \bibinfo{pages}{043001} (\bibinfo{year}{2001}).

\bibitem[{\citenamefont{Novikov et~al.}(2000)\citenamefont{Novikov, Schmalzing,
  and Mukhanov}}]{novikov}
\bibinfo{author}{\bibfnamefont{D.}~\bibnamefont{Novikov}},
  \bibinfo{author}{\bibfnamefont{J.}~\bibnamefont{Schmalzing}},
  \bibnamefont{and} \bibinfo{author}{\bibfnamefont{V.~F.}
  \bibnamefont{Mukhanov}}, \bibinfo{journal}{A\&A}
  \textbf{\bibinfo{volume}{364}}, \bibinfo{pages}{17} (\bibinfo{year}{2000}).

\bibitem[{\citenamefont{Gott et~al.}(1990)}]{gott}
\bibinfo{author}{\bibfnamefont{J.~R.} \bibnamefont{Gott}} \bibnamefont{et~al.},
  \bibinfo{journal}{ApJ} \textbf{\bibinfo{volume}{352}}, \bibinfo{pages}{1}
  (\bibinfo{year}{1990}).

\bibitem[{\citenamefont{Komatsu et~al.}(2003)}]{WMAPKomatsu}
\bibinfo{author}{\bibfnamefont{E.}~\bibnamefont{Komatsu}} \bibnamefont{et~al.},
  \bibinfo{journal}{ApJS} \textbf{\bibinfo{volume}{148}}, \bibinfo{pages}{119}
  (\bibinfo{year}{2003}).

\bibitem[{\citenamefont{Eriksen et~al.}(2002)\citenamefont{Eriksen, Banday, and
  G\'orski}}]{eriksen}
\bibinfo{author}{\bibfnamefont{H.~K.} \bibnamefont{Eriksen}},
  \bibinfo{author}{\bibfnamefont{A.~J.} \bibnamefont{Banday}},
  \bibnamefont{and} \bibinfo{author}{\bibfnamefont{K.~M.}
  \bibnamefont{G\'orski}}, \bibinfo{journal}{A\&A}
  \textbf{\bibinfo{volume}{395}}, \bibinfo{pages}{409} (\bibinfo{year}{2002}).

\bibitem[{\citenamefont{Heavens and Gupta}(2000)}]{heavens}
\bibinfo{author}{\bibfnamefont{A.~F.} \bibnamefont{Heavens}} \bibnamefont{and}
  \bibinfo{author}{\bibfnamefont{S.}~\bibnamefont{Gupta}},
  \bibinfo{journal}{MNRAS} \textbf{\bibinfo{volume}{324}}, \bibinfo{pages}{960}
  (\bibinfo{year}{2000}).

\bibitem[{\citenamefont{Scaramella and Vittorio}(1991)}]{vittorio}
\bibinfo{author}{\bibfnamefont{R.}~\bibnamefont{Scaramella}} \bibnamefont{and}
  \bibinfo{author}{\bibfnamefont{N.}~\bibnamefont{Vittorio}},
  \bibinfo{journal}{ApJ} \textbf{\bibinfo{volume}{375}}, \bibinfo{pages}{439}
  (\bibinfo{year}{1991}).

\bibitem[{\citenamefont{Dor\'e et~al.}(2003)}]{dore}
\bibinfo{author}{\bibfnamefont{O.}~\bibnamefont{Dor\'e}}, \bibinfo{author}
   {\bibnamefont{S.~Colombi}} \bibnamefont{and} \bibinfo{author}{\bibfnamefont{F.~R.}~\bibfnamefont{Bouchet}}
  \bibinfo{journal}{MNRAS} \textbf{\bibinfo{volume}{344}}, \bibinfo{pages}{905}
  (\bibinfo{year}{2003}).
	

\bibitem[{\citenamefont{Barreiro et~al.}(2001)\citenamefont{Barreiro,
  Martinez-Gonzalez, and Sanz}}]{barreiro2}
\bibinfo{author}{\bibfnamefont{R.~B.} \bibnamefont{Barreiro}},
  \bibinfo{author}{\bibfnamefont{E.}~\bibnamefont{Martinez-Gonzalez}},
  \bibnamefont{and} \bibinfo{author}{\bibfnamefont{J.~L.} \bibnamefont{Sanz}},
  \bibinfo{journal}{MNRAS} \textbf{\bibinfo{volume}{322}}, \bibinfo{pages}{411}
  (\bibinfo{year}{2001}).

\bibitem[{\citenamefont{Phillips and Kogut}(2000)}]{phillips}
\bibinfo{author}{\bibfnamefont{N.~G.} \bibnamefont{Phillips}} \bibnamefont{and}
  \bibinfo{author}{\bibfnamefont{A.}~\bibnamefont{Kogut}},
  \bibinfo{journal}{ApJ} \textbf{\bibinfo{volume}{548}}, \bibinfo{pages}{540}
  (\bibinfo{year}{2000}).

\bibitem[{\citenamefont{Komatsu et~al.}(2002)\citenamefont{Komatsu, Wandelt,
  Spergel, Banday, and G\'orski}}]{komatsu}
\bibinfo{author}{\bibfnamefont{E.}~\bibnamefont{Komatsu}},
  \bibinfo{author}{\bibfnamefont{B.~D.} \bibnamefont{Wandelt}},
  \bibinfo{author}{\bibfnamefont{D.~N.} \bibnamefont{Spergel}},
  \bibinfo{author}{\bibfnamefont{A.~J.} \bibnamefont{Banday}},
  \bibnamefont{and} \bibinfo{author}{\bibfnamefont{K.~M.}
  \bibnamefont{G\'orski}}, \bibinfo{journal}{ApJ}
  \textbf{\bibinfo{volume}{566}}, \bibinfo{pages}{19} (\bibinfo{year}{2002}).

\bibitem[{\citenamefont{Komatsu and Spergel}(2001)}]{komatsuspergel}
\bibinfo{author}{\bibfnamefont{E.}~\bibnamefont{Komatsu}} \bibnamefont{and}
  \bibinfo{author}{\bibfnamefont{D.~N.} \bibnamefont{Spergel}},
  \bibinfo{journal}{Phys Rev.} \textbf{\bibinfo{volume}{D63}},
  \bibinfo{pages}{063002} (\bibinfo{year}{2001}).

\bibitem[{\citenamefont{Troia et~al.}(2003)()}]{grazia}
\bibinfo{author}{\bibfnamefont{G.~D.} \bibnamefont{Troia}}
  \bibnamefont{et~al.}, \bibinfo{journal}{MNRAS} \textbf{\bibinfo{volume}{313}}, \bibinfo{pages}{284} (\bibinfo{year}{2003}).

\bibitem[{\citenamefont{Winitzki and Wu}()}]{win}
\bibinfo{author}{\bibfnamefont{S.}~\bibnamefont{Winitzki}} \bibnamefont{and}
  \bibinfo{author}{\bibfnamefont{J.~H.~P.} \bibnamefont{Wu}},
  \eprint{astro-ph/0007213}.

\bibitem[{\citenamefont{Hu}(2001)}]{hu}
\bibinfo{author}{\bibfnamefont{W.}~\bibnamefont{Hu}}, \bibinfo{journal}{Phys.
  Rev.} \textbf{\bibinfo{volume}{D64}}, \bibinfo{pages}{083005}
  (\bibinfo{year}{2001}).

\bibitem[{\citenamefont{Kunz et~al.}(2001)}]{kunz}
\bibinfo{author}{\bibfnamefont{M.}~\bibnamefont{Kunz}} \bibnamefont{et~al.},
  \bibinfo{journal}{ApJ} \textbf{\bibinfo{volume}{563}}, \bibinfo{pages}{L99}
  (\bibinfo{year}{2001}).

\bibitem[{\citenamefont{Chiang et~al.}()\citenamefont{Chiang, Naselsky, and
  Coles}}]{naselsky}
\bibinfo{author}{\bibfnamefont{L.~Y.} \bibnamefont{Chiang}},
  \bibinfo{author}{\bibfnamefont{P.}~\bibnamefont{Naselsky}}, \bibnamefont{and}
  \bibinfo{author}{\bibfnamefont{P.}~\bibnamefont{Coles}},
  \eprint{astro-ph/0208235}.

\bibitem[{\citenamefont{Wu et~al.}(2001)}]{boom}
\bibinfo{author}{\bibfnamefont{J.~H.} \bibnamefont{Wu}} \bibnamefont{et~al.},
  \bibinfo{journal}{Phys. Rev. Lett.} \textbf{\bibinfo{volume}{87}},
  \bibinfo{pages}{251303} (\bibinfo{year}{2001}).

\bibitem[{\citenamefont{Polenta et~al.}(2002)}]{polenta}
\bibinfo{author}{\bibfnamefont{G.}~\bibnamefont{Polenta}} \bibnamefont{et~al.},
  \bibinfo{journal}{ApJ Lett.} \textbf{\bibinfo{volume}{572}},
  \bibinfo{pages}{L27} (\bibinfo{year}{2002}).

\bibitem[{\citenamefont{Ferreira et~al.}(1998)\citenamefont{Ferreira, Magueijo,
  and G\'orski}}]{cobeng1}
\bibinfo{author}{\bibfnamefont{P.~G.} \bibnamefont{Ferreira}},
  \bibinfo{author}{\bibfnamefont{J.}~\bibnamefont{Magueijo}}, \bibnamefont{and}
  \bibinfo{author}{\bibfnamefont{K.~M.} \bibnamefont{G\'orski}},
  \bibinfo{journal}{ApJ} \textbf{\bibinfo{volume}{503}}, \bibinfo{pages}{L1}
  (\bibinfo{year}{1998}).

\bibitem[{\citenamefont{Banday et~al.}(2000)\citenamefont{Banday, Zaroubi, and
  G\'orski}}]{cobeng2}
\bibinfo{author}{\bibfnamefont{A.~J.} \bibnamefont{Banday}},
  \bibinfo{author}{\bibfnamefont{S.}~\bibnamefont{Zaroubi}}, \bibnamefont{and}
  \bibinfo{author}{\bibfnamefont{K.~M.} \bibnamefont{G\'orski}},
  \bibinfo{journal}{ApJ} \textbf{\bibinfo{volume}{533}}, \bibinfo{pages}{575}
  (\bibinfo{year}{2000}).

\bibitem[{\citenamefont{Hansen et~al.}(2002)\citenamefont{Hansen, Marinucci,
  Natoli, and Vittorio}}]{paper1}
\bibinfo{author}{\bibfnamefont{F.~K.} \bibnamefont{Hansen}},
  \bibinfo{author}{\bibfnamefont{D.}~\bibnamefont{Marinucci}},
  \bibinfo{author}{\bibfnamefont{P.}~\bibnamefont{Natoli}}, \bibnamefont{and}
  \bibinfo{author}{\bibfnamefont{N.}~\bibnamefont{Vittorio}},
  \bibinfo{journal}{Phys. Rev.} \textbf{\bibinfo{volume}{D66}},
  \bibinfo{pages}{063006} (\bibinfo{year}{2002}).

\bibitem[{\citenamefont{Hansen et~al.}(2003)\citenamefont{Hansen, Marinucci,
  and Vittorio}}]{paper2}
\bibinfo{author}{\bibfnamefont{F.~K.} \bibnamefont{Hansen}},
  \bibinfo{author}{\bibfnamefont{D.}~\bibnamefont{Marinucci}},
  \bibnamefont{and} \bibinfo{author}{\bibfnamefont{N.}~\bibnamefont{Vittorio}},
  \bibinfo{journal}{Phys. Rev.} \textbf{\bibinfo{volume}{D67}},
  \bibinfo{pages}{123004} (\bibinfo{year}{2003}).

\bibitem[{\citenamefont{Marinucci and Piccioni}(2004)}]{mar}
\bibinfo{author}{\bibnamefont{D.}~\bibnamefont{Marinucci}} \bibnamefont{and}
  \bibinfo{author}{\bibnamefont{M.}~\bibnamefont{Piccioni}}, 
\bibinfo{journal}{Ann. Stat.} \textbf{\bibinfo{volume}{32}},
  \bibinfo{pages}{3} (\bibinfo{year}{2004}).

\bibitem[{\citenamefont{Barreiro and Hobson}(2001)}]{barreiro1}
\bibinfo{author}{\bibfnamefont{R.~B.} \bibnamefont{Barreiro}} \bibnamefont{and}
  \bibinfo{author}{\bibfnamefont{M.~P.} \bibnamefont{Hobson}},
  \bibinfo{journal}{MNRAS} \textbf{\bibinfo{volume}{327}}, \bibinfo{pages}{813}
  (\bibinfo{year}{2001}).

\bibitem[{\citenamefont{Barreiro et~al.}(2000)\citenamefont{Barreiro, Hobson,
  Lasenby, Banday, Gorski, and Hinshaw}}]{barreiro3}
\bibinfo{author}{\bibfnamefont{R.~B.} \bibnamefont{Barreiro}},
  \bibinfo{author}{\bibfnamefont{M.~P.} \bibnamefont{Hobson}},
  \bibinfo{author}{\bibfnamefont{A.~N.} \bibnamefont{Lasenby}},
  \bibinfo{author}{\bibfnamefont{A.~J.} \bibnamefont{Banday}},
  \bibinfo{author}{\bibfnamefont{K.~M.} \bibnamefont{Gorski}},
  \bibnamefont{and} \bibinfo{author}{\bibfnamefont{G.}~\bibnamefont{Hinshaw}},
  \bibinfo{journal}{MNRAS} \textbf{\bibinfo{volume}{318}}, \bibinfo{pages}{475}
  (\bibinfo{year}{2000}).

\bibitem[{\citenamefont{Gonzalez et~al.}(2002)\citenamefont{Gonzalez, Gallegos,
  Argueso, Cayon, and Sanz}}]{mg}
\bibinfo{author}{\bibfnamefont{E.~M.} \bibnamefont{Gonzalez}},
  \bibinfo{author}{\bibfnamefont{J.~E.} \bibnamefont{Gallegos}},
  \bibinfo{author}{\bibfnamefont{F.}~\bibnamefont{Argueso}},
  \bibinfo{author}{\bibfnamefont{L.}~\bibnamefont{Cayon}}, \bibnamefont{and}
  \bibinfo{author}{\bibfnamefont{J.~L.} \bibnamefont{Sanz}},
  \bibinfo{journal}{MNRAS} \textbf{\bibinfo{volume}{336}}, \bibinfo{pages}{22}
  (\bibinfo{year}{2002}).

\bibitem[{\citenamefont{Mukherjee et~al.}(2000)\citenamefont{Mukherjee, Hobson,
  and Lasenby}}]{cobewav}
\bibinfo{author}{\bibfnamefont{P.}~\bibnamefont{Mukherjee}},
  \bibinfo{author}{\bibfnamefont{M.~P.} \bibnamefont{Hobson}},
  \bibnamefont{and} \bibinfo{author}{\bibfnamefont{A.~N.}
  \bibnamefont{Lasenby}}, \bibinfo{journal}{MNRAS}
  \textbf{\bibinfo{volume}{318}}, \bibinfo{pages}{1157} (\bibinfo{year}{2000}).

\bibitem[{\citenamefont{Minkowski}(1903)}]{Minkowski}
\bibinfo{author}{\bibfnamefont{H.}~\bibnamefont{Minkowski}},
  \bibinfo{journal}{Math. Ann.} \textbf{\bibinfo{volume}{57}},
  \bibinfo{pages}{447} (\bibinfo{year}{1903}).

\bibitem[{\citenamefont{Worsley}(1994)}]{Worsley}
\bibinfo{author}{\bibfnamefont{K.~J.} \bibnamefont{Worsley}},
  \bibinfo{journal}{Adv. Appl. Prob.} \textbf{\bibinfo{volume}{26}},
  \bibinfo{pages}{13} (\bibinfo{year}{1994}).

\bibitem[{\citenamefont{Schmalzing and G\'orski}(1998)}]{Schmalzing}
\bibinfo{author}{\bibfnamefont{J.}~\bibnamefont{Schmalzing}} \bibnamefont{and}
  \bibinfo{author}{\bibfnamefont{K.~M.} \bibnamefont{G\'orski}},
  \bibinfo{journal}{MNRAS} \textbf{\bibinfo{volume}{297}}, \bibinfo{pages}{355}
  (\bibinfo{year}{1998}).

\bibitem[{\citenamefont{Tomita}(1990)}]{Tomita}
\bibinfo{author}{\bibfnamefont{H.}~\bibnamefont{Tomita}},
  \bibinfo{journal}{Formation, Dynamics and Statistics of patterns, ed. K.
  Kawasaki, M. Suzuki, A. Onuki, Vol 1 (World Scientific), 113-157}
  (\bibinfo{year}{1990}).

\bibitem[{\citenamefont{G\'orski et~al.}(1998)\citenamefont{G\'orski, Hivon,
  and Wandelt}}]{healpix}
\bibinfo{author}{\bibfnamefont{K.~M.} \bibnamefont{G\'orski}},
  \bibinfo{author}{\bibfnamefont{E.}~\bibnamefont{Hivon}}, \bibnamefont{and}
  \bibinfo{author}{\bibfnamefont{B.~D.} \bibnamefont{Wandelt}},
  \bibinfo{journal}{Analysis Issues for Large CMB Data Sets', 1998, eds A. J.
  Banday, R. K. Sheth and L. Da Costa, ESO, Printpartners Ipskamp, NL,
  pp.37-42}  (\bibinfo{year}{1998}).

\bibitem[{\citenamefont{Pierpaoli}(2003)}]{sorgenti}
\bibinfo{author}{\bibfnamefont{E.}~\bibnamefont{Pierpaoli}},
  \bibinfo{journal}{ApJ} \textbf{\bibinfo{volume}{589}}, \bibinfo{pages}{58}
  (\bibinfo{year}{2003}).

\bibitem[{\citenamefont{Tegmark and de~Oliveira-Costa}(1998)}]{tegmark}
\bibinfo{author}{\bibfnamefont{M.}~\bibnamefont{Tegmark}} \bibnamefont{and}
  \bibinfo{author}{\bibfnamefont{A.}~\bibnamefont{de~Oliveira-Costa}},
  \bibinfo{journal}{ApJ} \textbf{\bibinfo{volume}{500}}, \bibinfo{pages}{L83}
  (\bibinfo{year}{1998}).

\bibitem[{\citenamefont{Schmalzing et~al.}(2000)\citenamefont{Schmalzing,
  Takada, and Futamase}}]{Schmalzing2}
\bibinfo{author}{\bibfnamefont{J.}~\bibnamefont{Schmalzing}},
  \bibinfo{author}{\bibfnamefont{M.}~\bibnamefont{Takada}}, \bibnamefont{and}
  \bibinfo{author}{\bibfnamefont{T.}~\bibnamefont{Futamase}},
  \bibinfo{journal}{ApJ} \textbf{\bibinfo{volume}{544}}, \bibinfo{pages}{L83}
  (\bibinfo{year}{2000}).

\bibitem[{\citenamefont{Cay\'on et~al.}(2000)}]{points}
\bibinfo{author}{\bibfnamefont{L.}~\bibnamefont{Cay\'on}} \bibnamefont{et~al.},
  \bibinfo{journal}{MNRAS} \textbf{\bibinfo{volume}{315}}, \bibinfo{pages}{757}
  (\bibinfo{year}{2000}).

\end{thebibliography}

\end{document}